\documentclass[twocolumn]{aastex701}

\newcommand{\wprp}{\omega_\mathrm{p}(r_\mathrm{p})}
\newcommand{\Mprp}{M_\mathrm{p}(r_\mathrm{p})}
\newcommand{\rp}{r_\mathrm{p}}
\newcommand{\hMpc}{h^{-1} \mathrm{Mpc}}
\newcommand{\mass}{\log_{10}\left(M_{\star}/\mathrm{M}_{\sun}\right)}

\usepackage{multirow}
\usepackage{amsmath}

\begin{document}

\title{Environmental dependence of galaxy properties in the southern GAMA regions}

\correspondingauthor{Koti Joy}

\author[orcid=0009-0000-6201-5601]{Koti Joy}
\affiliation{Department of Earth and Space Sciences, Indian Institute of Space Science \& Technology, Thiruvananthapuram 695547, Kerala, India}
\affiliation{U R Rao Satellite Centre, Indian Space Research Organisation, Bengaluru 560017, Karnataka, India}
\email[show]{koti.joy95@gmail.com} 

\author[orcid=0000-0002-2210-0681,sname='US']{Unnikrishnan Sureshkumar}
\affiliation{National Centre for Nuclear Research, ul. Pasteura 7, 02-093 Warsaw, Poland}
\affiliation{Wits Centre for Astrophysics, School of Physics, University of the Witwatersrand, Private Bag 3, Johannesburg 2050, South Africa}
\email[show]{unni.suresh@ncbj.gov.pl}
    
\author[orcid=0000-0002-1490-5367,sname='AN']{Anand Narayanan}
\affiliation{Department of Earth and Space Sciences, Indian Institute of Space Science \& Technology, Thiruvananthapuram 695547, Kerala, India}
\email{anand@iist.ac.in}

\author[orcid=0000-0002-3818-8315,sname='AD']{Anna Durkalec}
\affiliation{National Centre for Nuclear Research, ul. Pasteura 7, 02-093 Warsaw, Poland}
\email{Anna.Durkalec@ncbj.gov.pl}

\author[orcid=0000-0003-4169-9738,sname='SB']{Sabine Bellstedt}
\affiliation{ICRAR, The University of Western Australia, 7 Fairway, Crawley WA 6009, Australia}
\email{sabine.bellstedt@uwa.edu.au}

\author[orcid=0000-0003-3358-0665,sname='AP']{Agnieszka Pollo}
\affiliation{National Centre for Nuclear Research, ul. Pasteura 7, 02-093 Warsaw, Poland}
\affiliation{Astronomical Observatory of the Jagiellonian University, ul. Orla 171, 30-244 Kraków, Poland}
\email{Agnieszka.Pollo@ncbj.gov.pl}

\author[orcid=0000-0002-8490-8117,sname='MH']{Matt Hilton} 
\affiliation{Wits Centre for Astrophysics, School of Physics, University of the Witwatersrand, Private Bag 3, Johannesburg 2050, South Africa}
\affiliation{Astrophysics Research Centre, School of Mathematics, Statistics, and Computer Science, University of KwaZulu-Natal, Westville Campus, Durban 4041, South Africa}
\email{matt.hilton@wits.ac.za}


\begin{abstract}

Using data from the Galaxy and Mass Assembly (GAMA) survey, we investigate how galaxy properties correlate with the local environment, focusing on its two southern regions (G02 and G23) that have not previously been examined in this context.
We employ two-point and marked correlation functions to quantify the environmental dependence of galaxy color, stellar mass, luminosity across the $u$, $g$, $r$, $J$, and $K$ bands, as well as star formation rate (SFR) and specific star formation rate (sSFR).
Our results show that $u-r$ and $g-r$ colors are most strongly correlated with local overdensity, followed by stellar mass. 
The SFR and sSFR exhibit a clear inverse relationship with environmental density, consistent with the trend observed for $u$-band luminosity, which traces young stellar populations.
The galaxies brighter in the $g$, $J$, and $K$ bands preferentially inhabit denser regions.
We interpret the environmental dependence of color using Halo Occupation Distribution (HOD) modeling.
The HOD modeling indicates that red galaxies trace dense environments more strongly than blue galaxies, exhibiting significantly higher satellite fractions and occupying more massive halos.
We demonstrate that both redshift and stellar mass incompleteness systematically affect clustering measurements, generally enhancing the observed environmental dependence.
Comparing our measurements from the southern regions with those from the equatorial GAMA regions, we find that the overall environmental trends remain consistent.
The measured correlations provide key constraints for models of galaxy assembly across mass and environment, while environmental trends in color and near-infrared luminosity trace stellar mass growth and quenching with redshift.

\end{abstract}



\section{Introduction} \label{sec:intro}

Galaxies live in various environments within the large-scale structure (LSS) of the universe, namely nodes, walls, filaments, and voids \citep{deLapparent1986,Bond1996,Alpaslan2014}.
The denser environments of the LSS, such as nodes and filaments, are occupied by redder, massive, more luminous and early-type galaxies, whereas less dense environments are dominated by bluer, less massive, less luminous, and late-type galaxies \citep{Dressler1980, zehavi2002ApJ...571..172Z, Blanton2005, guo2015MNRAS.453.4368G}.
These observations signify a strong correlation between the physical properties of the galaxies and their local environment.

The environments in which galaxies inhabit play a crucial role in shaping their evolution, determining key properties such as morphology, star formation rate (SFR), and luminosity \citep{kauffmann2004, peng2010}.
In denser environments galaxies are more likely to undergo interactions, including mergers and ram-pressure stripping \citep{GunnGott1972}, which can quench star formation and drive the transformation from star-forming spirals to quiescent ellipticals \citep{Oemler1974,DavisGeller1976,Dressler1980,Whitmore1993,Fasano2015}.
Galaxies in less dense environments, on the other hand, tend to retain their gas and continue star formation for extended periods.
Therefore, detailed studies on the environmental dependence of galaxy properties are essential for a better understanding of the role of environment in galaxy evolution

In most studies of galaxy property -- environment correlations, the environment is parameterized by a local density defined at a particular distance scale \citep[e.g.][]{muldrew2012}. However, it is important to examine these correlations over a wide range of scales to enable a more comprehensive investigation of environmental effects that operate at different scales, while minimizing the impact of the chosen density parameter.
In this work, we adapt such an approach by using galaxy clustering measurements such as the galaxy two-point correlation function \citep[2pCF;][]{Peebles1980} and marked correlation function \citep[MCF;][]{BK2000}

The 2pCF quantifies galaxy clustering as the excess probability of finding pairs of galaxies separated by a specific distance, compared to a random distribution, enabling a proper comparison of clustering strength between different samples.
Previous 2pCF measurements investigated how galaxy clustering depends on properties such as
color \citep{Zehavi2005,Zehavi2011,Skibba2014},
luminosity \citep{delaTorre2007,Zehavi2011,Norberg2001},
stellar mass \citep{Durkalec2018,Skibba2015}, and
SFR \citep{Hartley2010}.
Overall, these studies suggest that clustering strength is more pronounced for redder, passive galaxies with high stellar masses compared to late-type galaxies that are more actively star-forming.

Extending the 2pCF, the MCF incorporates additional information by weighting galaxies using specific properties, referred to as \textit{marks}.
The MCF measures the clustering of galaxies as a function of their properties, offering a refined understanding of how these properties are correlated with the local environment of the galaxies \citep{Sheth2005}.
Previous MCF studies have shown that galaxies that occupy the denser environments tend to be more luminous, redder, and older than those in less-dense environments \citep[e.g.][]{sheth2005_galform_models, sheth2006, Skibba2006, Skibba2009, Skibba2014, Unni_GAMA}.

\citealt{sheth2005_galform_models} applied marked statistics to semi-analytic models of galaxy formation and found that galaxies occurring in close pairs tend to be redder, more massive and exhibit on average, lower SFR.
Using age as mark, \citealt{sheth2006} showed that close galaxy pairs tend to host stellar populations that are older than average.
Similarly, MCFs were used to show that denser regions of the LSS are populated by galaxies that are brighter in optical bands \citep{Skibba2006}, early-type \citep{Skibba2009}, and redder \citep{Skibba2014}.
The MCFs have been demonstrated to be effective at identifying the property most strongly correlated with the environment.
For example in the local Universe, \citet{Unni_GAMA} found that stellar mass is the strongest tracer of the environment, followed by $K-$band luminosity.
At the same time, \cite{mons2025} observed that ultraviolet magnitude and color are strongly correlated with the environment at high redshifts.

The MCF has a significant advantage over the 2pCF.
Unlike the 2pCF, which requires the sample to be divided into distinct subsamples based on these properties, the MCF can make use of the full sample statistics to probe the scale-dependent environmental correlations of galaxy properties.
The MCFs also facilitate a direct comparison between various properties in terms of the environment \citep{Skibba2013}.
The refinement offered by the MCF allows for a more detailed examination of galaxy properties in relation to their environments, proving to be particularly insightful in cases where the standard 2pCF may not capture the subtleties of galaxy clustering on smaller scales.

We apply the 2pCF and MCF analysis to a set of galaxy samples from the southern sky regions of the Galaxy and Mass Assembly \citep[GAMA;][]{Driver2009, driver2011} survey.
The GAMA survey has previously been used to study various aspects of environmental dependence \citep[e.g.][]{Wijesinghe2012,Brough2013,Alpaslan2015,Poudel2016,Wang2018}.
In particular, MCFs have been applied to the GAMA survey to study the environmental correlations of galaxy properties in optical \citep{Gunawardhana2018, Unni_GAMA} and infrared \citep[IR;][]{Unni2023} bands, and galaxy merger probabilities \citep{Unni2024}, as well as that of galaxy groups \citep{Riggs2021}.

All the aforementioned environmental studies focus on the equatorial regions (G09, G12, G15) of the GAMA survey.
In this work, we focus on the southern regions.
By doing so, we build a complete picture of correlations between galaxy properties and the environment as seen through the whole GAMA survey.
Additionally, this allows us to explore cosmic variance effects.
Building upon established work on cosmic variance in the GAMA regions \citep{DriverRobotham2010,Driver2022}, we explore whether it significantly affects the measured environmental dependence of galaxy properties by comparing the MCFs from southern regions in this work with those from the equatorial regions presented in \citet{Unni_GAMA}.
The galaxy properties that we investigate for environmental dependance are the $u-r$, $g-r$ observed frame colors, $u$, $g$, $r$, $J$, and $K$ band luminosities, stellar mass, SFR, and sSFR (specific SFR).
We also interpret the observed color dependence of galaxy clustering within the Halo Occupation Distribution (HOD) framework to investigate the galaxy–halo connection.
Additionally, we study how stellar mass and redshift incompleteness affects clustering measurements.
Our results show that while cosmic variance can affect the amplitudes of the MCFs, the overall environmental trends remain the same.

The paper is structured as follows.
We describe the data and sample selection methods  in Section~\ref{sec:data}.
The methodology and error estimation methods are described in Section~\ref{sec:methodology}.
We present our results in Section~\ref{sec:results} and discuss them in Section~\ref{sec:discussion}.
The key results are summarized in Section~\ref{sec:conclusion}. This work adopts a flat $\Lambda$CDM cosmological model characterized by matter density $\Omega_\mathrm{M}=0.3$ and dark energy density, $\Omega_\Lambda=0.7$.
The Hubble constant is parametrized via $h = H_0/100\ \mathrm{km\ s^{-1}\ Mpc^{-1}}$, with galaxy properties measured using $h = 0.7$, except for those from the ProSpect catalogue which were measured using $h = 0.678,\ \Omega_\mathrm{M}=0.308,\ \Omega_\Lambda=0.692$ \citep{Bellstedt2020b}. 
All distances are given in comoving coordinates and are expressed in units of $\hMpc$.

\section{Data} \label{sec:data}

The Galaxy And Mass Assembly (GAMA; \citealt{driver2011}) survey is a multi-wavelength spectroscopic survey that spans from the ultraviolet to the far-IR range (0.15 -- 500 $\mu$m) using 21 broadband filters.
The survey covers five regions on the sky, namely the equatorial regions of G09, G12, and G15 and the southern regions of G02 and G23.
The survey aims to study galaxy evolution and test the $\Lambda$CDM model of structure formation by combining data from various ground-based and space-based facilities.
We refer the reader to \cite{Driver2009} for a detailed description of the GAMA survey, whereas below we briefly provide the details that are relevant to this work.


In this work, we used galaxies from the southern (G02 and G23) regions of the GAMA data release 4 \citep{Driver2022}. 
In all our samples, we select galaxies with secure spectroscopic redshifts using the flag \texttt{nQ}~$\geq$~3 \citep{Liske2015} and within the redshift limits of $0.1~<~z~<~0.2$.

\subsection{Region G02}\label{subsec:data_g02}

The G02 region covers a survey area of $55.7\ \mathrm{deg}^2$ with RA ranging from $30.2\degr$ to $38.8\degr$ and DEC ranging from $-10.25\degr$ to $-3.72\degr$.
The G02 region overlaps significantly with the Canada–France–Hawaii Telescope
Legacy Survey (CFHTLS)-W1 field \citep{gwyn2012_cfhtls} and Sloan Digital Sky Survey \citep[SDSS;][]{york2000_sdss_gen}, encompassing around 87\% of the former survey region.
Additionally, the X-ray Multi-Mirror eXtra Large \citep[XXL;][]{pierre2016_xxl} survey covers approximately 25 deg$^2$ of the G02 region.
The imaging data from CFHTLS and SDSS were used to define the target selection in the G02 region.
The high spectroscopic completeness area within G02 was determined based on the coverage of the XXL survey \citep{Baldry2018}.

The G02 region has varying levels of redshift completeness.
The redshift completeness is defined as the percentage of targets with reliable redshifts (\texttt{nQ~$\geq$~3}).
For the main survey area north of $-6.0\degr$ in DEC, the redshift completeness is 95.5\%.
The completeness drops to 46.4\% between $-6.3\degr$ and $-6.0\degr$, and to 31.0\% on average to the south of $-6.3\degr$.

We use the catalogue \textsc{G02TilingCatv07} \citep{Baldry2018} to select the SDSS selected galaxies (\texttt{SURVEY\_CLASS} $\geq$ 5). 
The catalogue \textsc{G02SDSSInputCatv01} \citep{Baldry2018} provides astrometry and apparent magnitudes (to calculate observed frame colors) from SDSS and \textsc{StellarMassesG02SDSSv24} catalogue \citep{taylor2011_gama_stellar_mass} provides stellar masses and absolute magnitudes obtained using the SDSS photometry.

\begin{figure*}
    \centering
    \includegraphics[width=\linewidth]{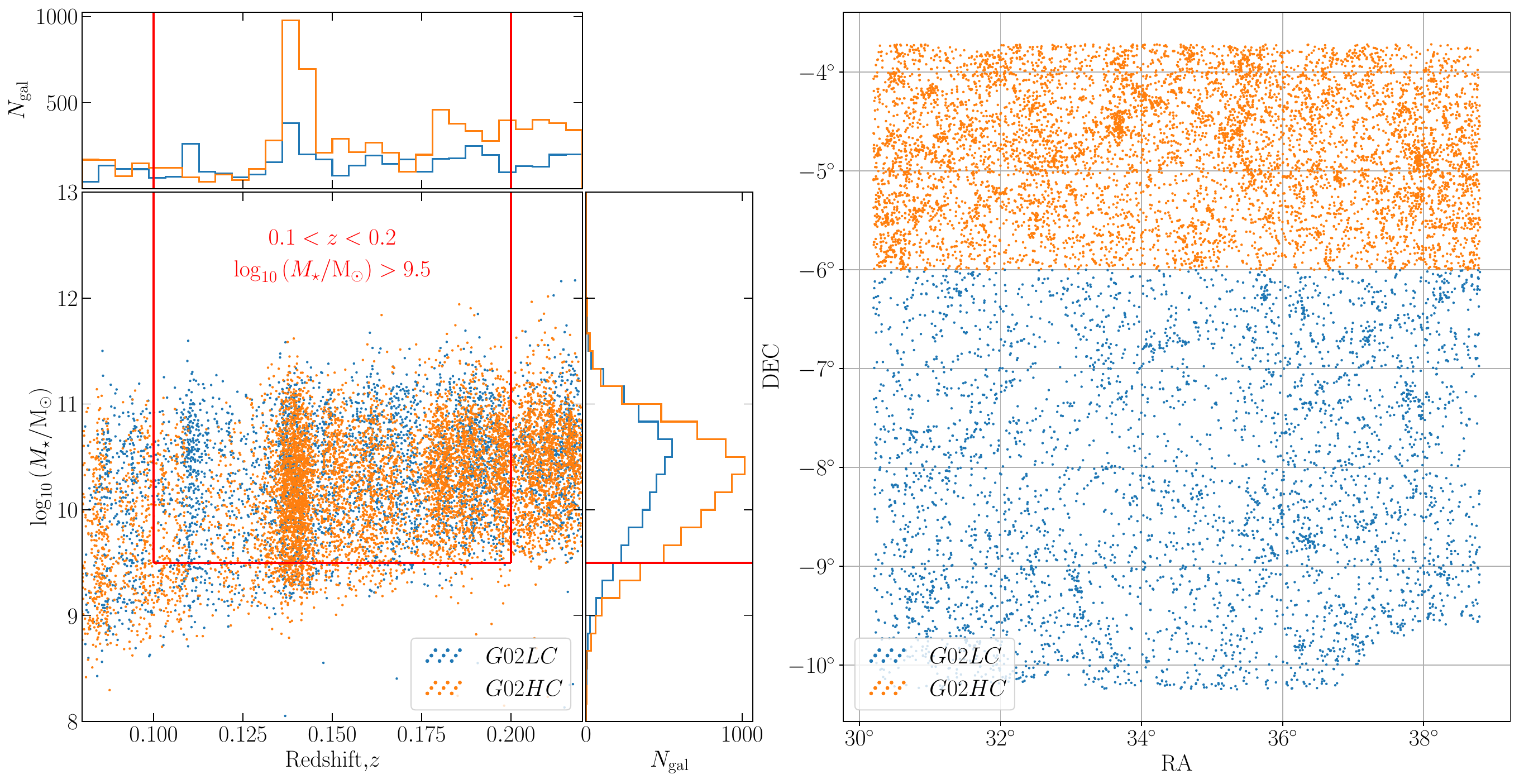}
    \caption{
    \textit{Left panel}: Selection of samples from the G02 region.
    The red lines indicate a redshift limit from 0.1 to 0.2 and a stellar mass cut of $\mass > 9.5$. 
    Galaxies within the red lines form the samples $G02LC$ (blue dots) and $G02HC$ (orange dots).
    \textit{Right panel}: RA-DEC distribution of  galaxies in the selected samples $G02LC$ and $G02HC$.}
    \label{fig:Region_G02_mass_distribution}
\end{figure*}

We create two stellar-mass limited subsamples from the G02 region -- one covering the low complete region south of $-6.0\degr$ ($G02LC$) and the other covering the high-redshift completeness region north of $-6.0\degr$ ($G02HC$).
From both the subsamples, we select only galaxies above the stellar mass limit of $\log_{10} (M_{\star}/\mathrm{M}_{\odot})^\mathrm{min} = 9.5$.
This selection results in 5257 galaxies in $G02LC$ and 3029 in $G02HC$ samples.
The redshift~-~stellar mass distribution of both the samples, along with their RA-DEC distribution are shown in Fig~\ref{fig:Region_G02_mass_distribution}.



\subsection{Region G23}\label{subsec:data_g23}

\begin{figure*}
    \centering
    \includegraphics[width=\linewidth]{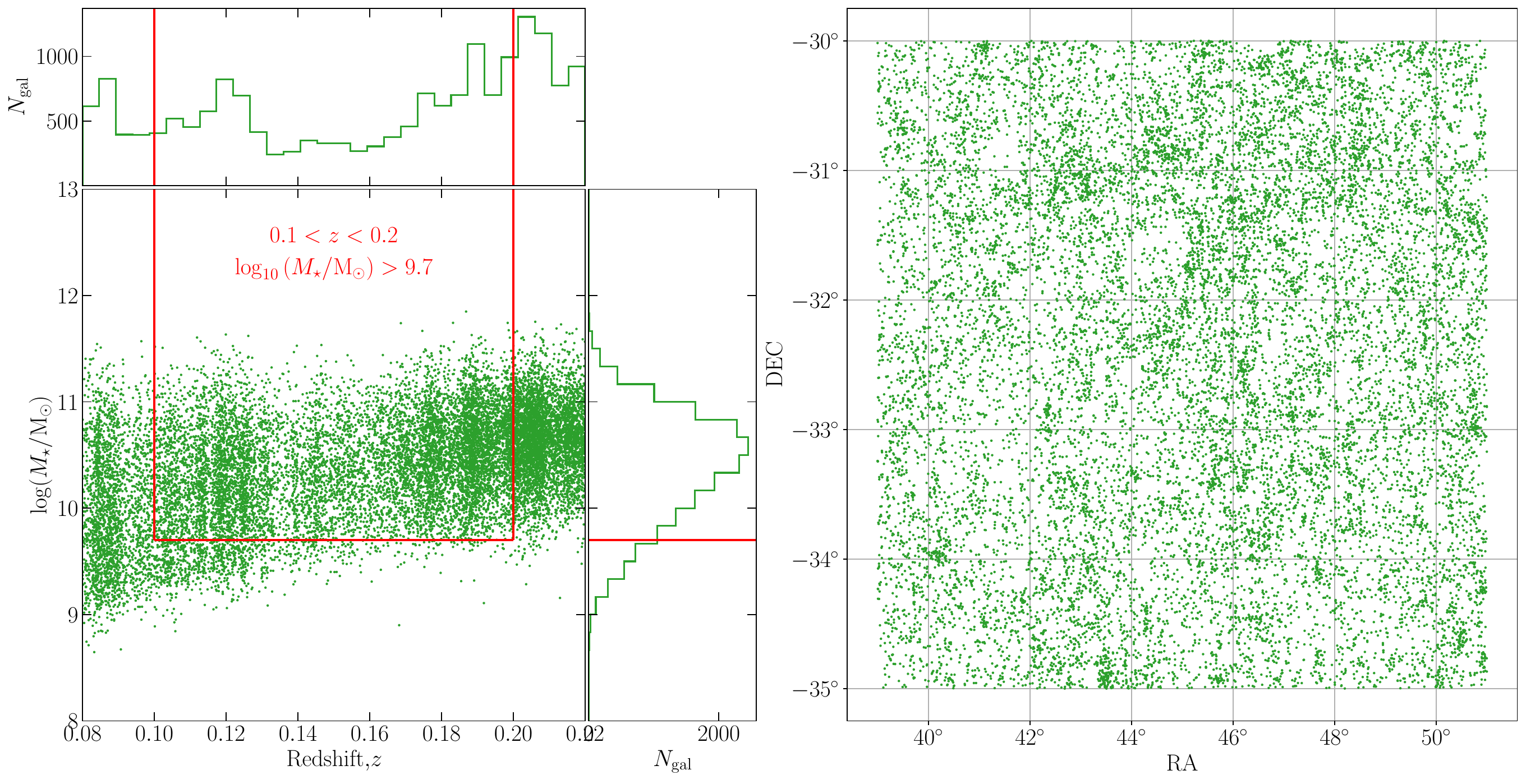}
    \caption{
    \textit{Left panel:} Selection of galaxy sample from the G23 region. 
    The red lines indicate a redshift limit from 0.1 to 0.2 and a stellar mass cut of $\mass > 9.7$.
    Galaxies within the red lines form the sample $G23$.
    \textit{Right panel}: RA-DEC distribution of galaxies in sample $G23$.}
    \label{fig:Region_G23_mass_distribution}
\end{figure*}

The G23 region covers an RA range of $339.0\degr$ to $351.0\degr$ and a DEC range of $-35\degr$ to $-30\degr$, covering a survey area of $50.6\ \mathrm{deg}^2$.
We select sample $G23$ using the catalogue \textsc{gkvScienceCatv02} \citep{Bellstedt2020a}, which provides astrometry and fluxes.
We convert the fluxes into apparent magnitudes and use these to calculate observed frame colors.
We select galaxies using the filter \texttt{uberclass} = 1 and with magnitudes $r<19.42$ which corresponds to 95\% redshift completeness in region G23 \citep{Driver2022}.
We use the catalogues \textsc{StellarMassesGKVv24} \citep{Driver2022} to get absolute magnitudes and \textsc{ProSpectv02} \citep{Bellstedt2020a} to get stellar masses and SFRs.
We select only those galaxies above the stellar mass limit of $\log_{10} (M_{\star}/\mathrm{M}_{\odot})^\mathrm{min} = 9.7$, yielding 9472 galaxies in the $G23$ sample.
The difference in stellar mass limits between this sample and those from region G02 is due to different mass completeness levels in the two regions.
The redshift~-~stellar mass distribution and the sky distribution of the sample are shown in Fig.~\ref{fig:Region_G23_mass_distribution}.



An overview of the samples $G02LC$, $G02HC$, and $G23$ is given in Table~\ref{tab:samples} and their properties are given in Table~\ref{tab:prop_samples}.
We further divide sample $G23$ into red and blue galaxies based on $g-r$ color, and their properties are given in Table~\ref{tab:prop_samples}.

\begin{deluxetable}{c c c c c}
    \tablecaption{\label{tab:samples}Definitions of all the samples used in this work in the redshift range of $0.1<z<0.2$.}
    \tablehead{
        \colhead{$\log_{10} \left(M_{\star}/\mathrm{M}_{\odot} \right)^\mathrm{min}$} &
        \colhead{Sample} &
        \colhead{$m_\mathrm{lim}$} &
        \colhead{$z_\mathrm{compl}$} &
        \colhead{$N_\mathrm{gal}$}
    }
    \startdata
    \multirow{2}{*}{9.5} & $G02HC$ & \multirow{2}{*}{19.80} & 95.5\% & 5257 \\
                         & $G02LC$ &                          & 31.6\% & 3029 \\
    \hline
    9.7                  & $G23$   & 19.42                    & 95.0\% & 9472 \\
    \enddata
    \tablecomments{The columns represent the stellar mass cut, sample name, main survey magnitude limit, redshift completeness, and number of galaxies.}
\end{deluxetable}

\subsection{Star formation rates}\label{subsec:data_sfr}

Along with absolute magnitudes and stellar masses, we use the SFRs and sSFRs of galaxies in the MCF analysis.
For sample $G23$, the SFRs are taken from the \textsc{ProSpectv02} catalogue \citep{Bellstedt2020b}, which was made using the spectral energy distribution (SED) fitting code \textsc{ProSpect} \citep{Robotham2020} applied to the catalogue \textsc{gkvInputCat}.
This method of SFR estimation uses \cite{BC2003} stellar population models along with \cite{Chabrier2003} initial mass function and the \cite{CF2000} dust attenuation law.

For sample $G02$, the \textsc{ProSpect}-based SFRs are not available and hence, we estimate the SFRs of all galaxies in the sample using Bayesian Analysis of Galaxies for Physical Inference and Parameter EStimation \citep[\textsc{Bagpipes};][]{Carnall2018} SED-fitting code.
The code uses \cite{BC2003} stellar population models along with \cite{KB2002} initial mass function and the \cite{Calzetti2000} dust attenuation law.

Although different SED fitting methods have a systematic impact on the resulting galaxy properties \citep{bellstedt2025}, we do not expect a significant propagated effect on the MCF measurements.
This is due to the implementation of rank-ordering MCF method, as detailed in Section.~\ref{subsec:meth_MCF}, which preserves the relative ordering of galaxies even if the absolute values of the property differ.
To confirm this, we perform a test comparing the MCFs using SFRs estimated with \textsc{Bagpipes} and \textsc{ProSpect} codes for all galaxies in the sample $G23$.
The results of this test are presented in Appendix.~\ref{app:comparision_SFRs}, and show that while the MCFs differ slightly at small scales, the overall trends remain consistent.


\begin{deluxetable*}{c c c c c c c c c c c}


    \tablecaption{\label{tab:prop_samples}Properties of the galaxy samples used in this work.}
    \tabletypesize{\scriptsize}
    \setlength{\tabcolsep}{2pt}
    \tablehead{
        \colhead{Sample} &
        \colhead{$\log(M_*/\mathrm{M_{\odot}})$} &
        \colhead{$M_u$} & 
        \colhead{$M_g$} &
        \colhead{$M_r$} &
        \colhead{$M_J$} &
        \colhead{$M_K$} &
        \colhead{$(u-r)$} &
        \colhead{$(g-r)$} &
        \colhead{\begin{tabular}[c]{@{}c@{}} SFR\\ $\mathrm{(M_{\sun}/yr)}$\end{tabular}} &
        \colhead{\begin{tabular}[c]{@{}c@{}}$\log$(sSFR)\end{tabular}}
    }

    \startdata
    $G02HC$ & $10.29^{+0.44}_{-0.47}$ & $-20.27^{+0.68}_{-0.81}$ & $-21.21^{+0.67}_{-0.80}$ & $-21.55^{+0.73}_{-0.84}$ & 
              ...               & ...                & $2.46^{+0.80}_{-0.76}$     & $0.90^{+0.24}_{-0.30}$   & $1.05^{+3.12}_{-1.00}$ & $-10.15^{+0.69}_{-2.24}$ \\
    $G02LC$ & $10.41^{+0.48}_{-0.54}$ & $-20.48^{+0.73}_{-0.79}$ & $-21.44^{+0.73}_{-0.84}$ & $-21.79^{+0.82}_{-0.89}$ & 
              ...               & ...                & $2.49^{+0.73}_{-0.77}$     & $0.91^{+0.24}_{-0.30}$   & $1.22^{+3.55}_{-1.20}$ & $-10.19^{+0.73}_{-2.37}$ \\
    $G23$ & $10.47^{+0.41}_{-0.45}$ & $-20.23^{+0.84}_{-0.88}$ & $-21.24^{+0.75}_{-0.85}$ & $-21.62^{+0.75}_{-0.86}$ & 
              $-21.98^{+0.87}_{-0.91}$  & $-21.77^{+0.92}_{-0.95}$ & $2.23^{+0.69}_{-0.59}$     & $0.83^{+0.25}_{-0.26}$   & $1.06^{+2.06}_{-1.06}$ & $-10.40^{+0.64}_{-3.39}$ \\ \hline
    $G23$ red & $10.70^{+0.32}_{-0.31}$ & $-20.12^{+0.89}_{-0.90}$ & $-21.33^{+0.81}_{-0.85}$ & $-21.84^{+0.77}_{-0.81}$ & 
              $-22.34^{+0.75}_{-0.79}$  & $-22.17^{+0.76}_{-0.81}$ & $2.71^{+0.38}_{-0.41}$     & $0.99^{+0.16}_{-0.13}$   & $0.23^{+1.54}_{-0.23}$ & $-11.30^{+0.88}_{-38.2}$\\
    $G23$ blue & $10.17^{+0.36}_{-0.30}$ & $-20.36^{+0.75}_{-0.85}$ & $-21.15^{+0.70}_{-0.83}$ & $-21.38^{+0.68}_{-0.82}$ & 
              $-21.54^{+0.69}_{-0.86}$  & $-21.27^{+0.72}_{-0.87}$ & $1.75^{+0.28}_{-0.28}$     & $0.63^{+0.11}_{-0.12}$   & $2.01^{+2.04}_{-1.09}$ & $-9.86^{+0.22}_{-0.38}$
    \enddata

    
    \tablecomments{For each galaxy sample, the table columns list the following properties: its designated label, the 16th, 50th (median), and 84th percentile values for $u$, $g$, $r$, $J$, and $K$ bands, $(u-r)$ and $(g-r)$ colors, SFR and sSFR.}
    
\end{deluxetable*}

\subsection{Random samples}\label{subsec:data_randoms}

The measurement of galaxy clustering involves a comparison of the spatial distribution of the real galaxies to a random galaxy distribution.
Therefore, we need a random galaxy catalogue that has the same sky coverage and redshift distribution as the real galaxy catalogue to measure the 2pCF.
To prevent shot noise from dominating on small scales, the random catalogue must be oversampled, ideally containing 5 to 10 times more galaxies than the real catalog.

Following \cite{Farrow2015}, we use the method of \cite{Cole2011} to generate random samples for all the real galaxy samples defined in Table~\ref{tab:samples}.
This method uses a maximum likelihood estimator to simultaneously determine both the luminosity function and the redshift-dependent overdensity.
Subsequently, the algorithm generates multiple clones of each galaxy from the real catalogue.
Each of these clones is then placed at a random redshift, drawn from a distribution that is consistent with the survey's accessible volume and magnitude constraints.

We used the publicly available Fortran95 subroutine\footnote{\url{http://astro.dur.ac.uk/~cole/publications.html\#software}} to create our random catalogues for all the samples in this work.
This subroutine outputs random redshifts.
We assign a random RA and DEC inside the sample's sky coverage area to each redshift. 
Additionally in the case of sample $G02LC$, we use a redshift completeness mask, as detailed in Appendix~\ref{app:completeness_mask}. 
Hence, the random galaxy samples have the same sky coverage area and same redshift distribution ($0.1<z<0.2$) of the observed galaxy sample.
Since the random catalogue is created by cloning the real galaxies, each random galaxy is assigned the stellar mass of its parent real galaxy.
The respective stellar mass limits of the real sample are then applied to the random sample as well.

\section{Methodology} \label{sec:methodology}

\subsection{The galaxy two-point correlation function}\label{subsec:meth_2pCF}

The galaxy 2pCF, $\xi(r)$ is a statistical tool that quantifies galaxy clustering.
It is defined as the excess probability of observing a pair of galaxies at a given comoving distance $r$ in a volume element $\mathrm{d}V$, compared to a random distribution \citep{Peebles1980}.
That is,
\begin{equation}\label{eq:2pCF}
    \mathrm{d}P = n^2[1+\xi(r)]\mathrm{d}V,
\end{equation}
where $n$ is the number density of galaxies.

To estimate the 2pCF, we use the \cite{LS1993}, estimator which is a widely adopted estimator due to its ability to minimize the effects arising due to finite survey volume and limited number of galaxies.
The estimator makes use of the count of data-data galaxy pairs $DD(r)$, data-random galaxy pairs $DR(r)$, and random-random galaxy pairs $RR(r)$.
At a given comoving separation $r$, $\xi(r)$ is estimated using,
\begin{equation}\label{eq:xi(r)}
    \xi(r) = \frac{\langle DD(r) \rangle - 2\langle DR(r) \rangle + \langle RR(r) \rangle}{\langle RR(r) \rangle},
\end{equation}
where $\langle \rangle$ refers to the quantity normalized by the total number of such pairs.

We measure the two-dimensional correlation function, $\xi(\rp,\pi)$, by resolving the comoving separation of galaxy pairs into two components: parallel ($\pi$) and perpendicular ($\rp$) to the observer's line of sight.
This method is necessary to isolate the distortions caused by peculiar velocities.
Subsequently, we compute the projected two-point correlation function, $\wprp$, by integrating with respect to $\pi$ up to a maximum separation, $\pi_\mathrm{max}$.
That is,
\begin{equation}\label{eq:omega(r_p)}
    \wprp = 2\int^{\pi_\mathrm{max}}_0 \xi(\rp,\pi)d\pi.
\end{equation}
The choice of $\pi_\mathrm{max}$ should be optimal, ensuring that all correlated pairs are included while minimizing the noise in the estimator.
Following previous clustering works in GAMA \citep[e.g.][]{loveday2018_gama_pvd, Farrow2015,Unni_GAMA}, we choose $\pi_\mathrm{max}$ to be $40\ \hMpc$.

\subsection{Marked correlation function}\label{subsec:meth_MCF}

While the 2pCF characterizes the spatial distribution of galaxies, the marked correlation function enables the study of the correlations between galaxy properties and their environment.
The process starts by assigning a mark to each galaxy. 
The mark can be a measure of any physical property of galaxies, such as luminosity, stellar mass, color, SFR and morphology.

The two-point MCF is defined as
\begin{equation} \label{eq:MCF}
    M(r) = \frac{1+W(r)}{1+\xi(r)},
\end{equation}
where $\xi(r)$ is the regular unweighted 2pCF and $W(r)$ is the weighted 2pCF.
The $W(r)$ is estimated using the weighted pair counts, with each galaxy in the pair weighted by the ratio of its mark to the mean value of that mark across the sample.

Similar to the 2pCF, we use projected two-point MCF, $M_\mathrm{p}(\rp)$, which is given by,
\begin{equation}\label{eq:M_p(r_p)}
    M_\mathrm{p}(\rp) = \frac{1+W_\mathrm{p}(\rp)/\rp}{1+\wprp/\rp}.
\end{equation}
%
The physical interpretation of the MCF at a scale $\rp$ is that it measures whether galaxies in pairs at that separation tend to have mark values that are, on average, enhanced or suppressed relative to the mean mark of the complete sample \citep{sheth&tormen2004}.
An MCF value greater than unity for a given property at a specific separation suggests a higher probability of finding galaxy pairs at that separation, with both galaxies in the pair exhibiting stronger values in that property.
Therefore, $\text{MCF} > 1$ indicates a stronger correlation between the property and the environment, while the opposite implies an anti-correlation.

To facilitate consistent comparisons of MCFs obtained using different galaxy properties, we employ the rank-ordering technique proposed by \cite{Skibba2013}.
Instead of using the property values directly as marks, we assign a rank to each galaxy based on the relative strength of its property within the sample.
That is, the galaxy with the highest luminosity (or any property considered) is assigned the highest rank, while the one with the lowest value receives the lowest rank.
These ranks, which have a uniform distribution across the sample, are then used as marks to weight the correlation function.
This rank-ordering approach ensures that the amplitudes of the MCFs measured using different properties are directly comparable, as they are all constructed from the same uniform distribution of ranks.

\subsection{Error estimates}\label{subsec:meth_err}

The uncertainties in 2pCF measurements are estimated using the jackknife resampling method \citep{Norberg2009}.
This method involves dividing the survey area into $N_\mathrm{JK}$ subregions and recomputing the 2pCF multiple times, each time omitting one subregion.
The scatter among these jackknife realizations is then used to construct the covariance matrix, which is given by,
\begin{equation}\label{eq:cov_matr}
    C_{ij} = \frac{N_\mathrm{JK}-1}{N_\mathrm{JK}}\sum^{N_\mathrm{JK}}_{k=1}\left(\omega_\mathrm{p}^k(r_i) - \bar{\omega}_\mathrm{p}(r_i)\right)\left(\omega_\mathrm{p}^k(r_j) - \bar{\omega}_\mathrm{p}(r_j)\right),
\end{equation}
where $\omega_\mathrm{p}^k(r_j)$ denotes the measurement of projected correlation function in the $j^\mathrm{th}$ $\rp$ bin from the $k^\mathrm{th}$ jackknife subsamples, with $\bar{\omega}_\mathrm{p}$ representing the average value over these subsamples.
The statistical errors on our measurements are subsequently taken from the square root of the diagonal terms of $C_{ij}$.
We use 24 jackknife samples in our computation so that $N_\mathrm{JK}>N_\mathrm{bin}^{3/2}$.
This ensures that the covariance matrix is not too noisy \citep{Mandelbaum2006}.
Based on \cite{Hartlap2007}, the inverse covariance matrix was corrected by the factor $(N_\mathrm{JK} - N_\mathrm{bin} - 2)/(N_\mathrm{JK} - 1)$. This corrected inverse covariance matrix was then used for the modeling of 2pCF.

The uncertainty in MCF has contributions from two components.
The first is spatial uncertainty, which arises due to the limited survey volume and is derived using jackknife resampling, similar to the approach used for the 2pCF.
The second component, mark uncertainty accounts for the correlation between galaxy positions and their assigned marks.
We quantify the mark uncertainty by repeatedly recalculating the MCF after randomly shuffling the marks among the galaxies in our sample.
The standard deviation across these multiple realizations serves as our estimate for this component of the error.
We combine these two components in quadrature to show the error bars in MCF.

\subsection{Power-law fitting}\label{subsec:meth_fitting}

The correlation function on the intermediate scales follows a power law \citep{GP1977}, and is given by
\begin{equation}\label{eq:xi_power}
    \xi(r) = \left(\frac{r}{r_0}\right)^{-\gamma},
\end{equation}
where $r_0$ and $\gamma$ are the correlation length and slope, respectively.

We fit for $r_0$ and $\gamma$ using the projected correlation function $\wprp$. 
Adopting the power-law parametrisation from Eq.~\ref{eq:xi_power}, allows the integral in Eq.~\ref{eq:omega(r_p)} to be evaluated analytically, which results to,
\begin{equation}\label{eq:power_law_omega_p}
    \omega_\mathrm{p} = \rp\left(\frac{\rp}{r_0}\right)^{-\gamma}\frac{\Gamma\left(\frac{1}{2}\right)\Gamma\left(\frac{\gamma-1}{2}\right)}{\Gamma\left(\frac{\gamma}{2}\right)}.
\end{equation}
We then minimize $\chi^2$, which is given by the following equation,
\begin{equation}\label{eq:chi2}
    \chi^2 = \sum_{i,j} \left(\omega_\mathrm{p}^\mathrm{mod}(r_i)-\omega_\mathrm{p}(r_i)\right) C_{ij}^{-1} \left(\omega_\mathrm{p}^\mathrm{mod}(r_j)-\omega_\mathrm{p}(r_j)\right) ,
\end{equation}
where $\omega_\mathrm{p}(r_i)$ represents the measured projected correlation function at $i^\mathrm{th}\ \rp$ bin, while $\omega_\mathrm{p}^\mathrm{mod}$ is the corresponding value predicted by the power-law model (Eq.~\ref{eq:power_law_omega_p}). The errors reported  for the best-fit parameters, $r_0$ and $\gamma$, correspond to the 68.3\% joint confidence levels.

\begin{deluxetable*}{c c c c c c c c c c c}


    \tablecaption{\label{tab:HODparams}Best-fit HOD parameters.}
    \tabletypesize{\scriptsize}
    \setlength{\tabcolsep}{2pt}
    \tablehead{
        \colhead{Sample} &
        \colhead{$\log (M_{\mathrm{min}} / h^{-1} \mathrm{M_{\odot}})$} &
        \colhead{$\log (M_{1} / h^{-1} \mathrm{M_{\odot}})$} & 
        \colhead{$\alpha$} &
        \colhead{$b_\mathrm{g}$} &
        \colhead{$\log (M_{\mathrm{eff}} / h^{-1} \mathrm{M_{\odot}})$} &
        \colhead{$f_\mathrm{s}$} &
        \colhead{$\chi^2/{\rm dof}$} &
    }

    \startdata
    $G02HC$ & $11.53^{+0.03}_{-0.03}$ & $12.96^{+0.04}_{-0.04}$ & $1.22^{+0.05}_{-0.06}$ & $1.15^{+0.03}_{-0.03}$ & 
              $13.45^{+0.04}_{-0.05}$ & $0.26^{+0.02}_{-0.02}$ & $0.98$ \\
    $G02LC$ & $12.08^{+0.03}_{-0.03}$ & $13.39^{+0.04}_{-0.04}$ & $1.25^{+0.05}_{-0.05}$ & $1.25^{+0.01}_{-0.01}$ & 
              $13.45^{+0.02}_{-0.02}$ & $0.25^{+0.03}_{-0.02}$ & $0.93$ \\
    $G23$ & $11.74^{+0.04}_{-0.04}$ & $13.07^{+0.05}_{-0.05}$ & $1.11^{+0.04}_{-0.04}$ & $1.13^{+0.01}_{-0.01}$ & 
              $13.33^{+0.02}_{-0.02}$ & $0.28^{+0.04}_{-0.03}$ & $1.91$ \\ \hline
    $G23$ red & $12.65_{-0.19}^{+0.53}$ & $14.04_{-0.28}^{+0.80}$ & $1.38_{-0.13}^{+0.19}$ & $1.46_{-0.06}^{+0.16}$ & $13.54_{-0.04}^{+0.03}$ & $0.13_{-0.21}^{+0.03}$ & 1.91 \\
    $G23$ blue & $11.96_{-0.03}^{+0.03}$ & $13.76_{-0.08}^{+0.06}$ & $1.40_{-0.13}^{+0.12}$ & $1.11_{-0.01}^{+0.01}$ & $13.23_{-0.04}^{+0.04}$ & $0.07_{-0.02}^{+0.02}$ & 0.57\\
    \enddata
    \tablecomments{The number of degrees of freedom is  6 for all samples corresponding to eight measured $\omega_\mathrm{p}(r_\mathrm{p})$ values and the number density, minus the three fitted HOD model parameters.}
\end{deluxetable*}

{\subsection{Halo Occupation Distribution modeling}\label{subsec:meth_HOD}
We also interpret the galaxy clustering within the Halo Occupation Distribution (HOD) formalism introduced by \cite{Zehavi2005}. In this framework, the mean occupations of galaxies is given by
\begin{equation}
\langle N_{\rm g} (M_\mathrm{h}) \rangle =
\begin{cases}
1 + \left(\dfrac{M_\mathrm{h}}{M_1}\right)^{\alpha}, & \text{for } M_\mathrm{h} > M_{\min}, \\
0, & \text{otherwise}.
\end{cases}
\end{equation}
where $M_\mathrm{h}$ denotes the dark matter halo mass. The mean halo occupation is the sum of the mean central and satellite occupation functions:
\begin{equation}
    \langle N_{\rm }(M_\mathrm{h})\rangle = \langle N_{\rm c}(M_\mathrm{h}) \rangle + \langle N_{\rm s}(M) \rangle
\end{equation}
The model contains three free parameters: the minimum halo mass $M_{\min}$ required for a halo to host a central galaxy above the adopted stellar-mass threshold, the characteristic satellite halo mass $M_1$, and the slope of the satellite occupation function $\alpha$.

We adopt the Navarro--Frenk--White density profile \citep{NFW1996}. The calculations are carried out using the publicly available \texttt{HALOMOD} package \citep{Murray2021}. 

The three-parameter HOD model used in this work provides a simple and robust description of galaxy clustering, and it has been adopted because the limitations of the measured projected correlation functions, with eight $r_\mathrm{p}$ bins spanning $0.1-10$ $h^{-1} \mathrm{Mpc}$. Introducing additional free parameters to HOD model would only lead to significant parameter degeneracies. However, this model simplicity comes with several limitations. In particular it assumes that galaxy occupation depends solely on halo mass, neglecting possible assembly biases, which has been shown to play important role in galaxy-halo connection \citep[e.g.,][]{Hearin2016, Hadzhiyska2021}. This may affect the absolute values of the best fitting parameters obtained in this work, but it is not expected to influence the qualitative differences between different correlation function fits.

The HOD parameters are constrained using a Markov Chain Monte Carlo method implemented with the \texttt{emcee} sampler \citep{Foreman-Mackey2013}. We use 25 walkers initialized at random positions in the parameter space. At each step, the likelihood is evaluated through the $\chi^2$ statistic:
\begin{equation}
\chi^2=
\left(
w_\mathrm{p}^{\rm mod}-w_\mathrm{p}^{\rm obs}
\right)^T
C^{-1}
\left(
w_\mathrm{p}^{\rm mod}-w_\mathrm{p}^{\rm obs}
\right)
+
\left(
\frac{
n_\mathrm{g}^{\rm mod}-n_\mathrm{g}^{\rm obs}
}{
\sigma_{n_\mathrm{g}}
}
\right)^2,
\end{equation}
where $w_\mathrm{p}^{\rm obs}$ and $w_\mathrm{p}^{\rm mod}$ denote the observed and model projected correlation functions, respectively. We incorporate the observed galaxy number density constraint $n_\mathrm{g}^\mathrm{obs}$, for which we assume a relative uncertainty of $5\%$.

The final parameter estimates are taken from the median of the marginalized posterior distributions, while the associated uncertainties correspond to the 16th and 84th percentiles.

The best-fitting HOD model allows us to derive the mean halo mass:
\begin{equation} \label{eq:effhalomass}
    \langle M_\mathrm{eff} \rangle =
    \int dM_\mathrm{h}\,
    M_\mathrm{h}\,n(M_\mathrm{h},z)
    \frac{\langle N_\mathrm{g}(M_\mathrm{h})\rangle}{n_\mathrm{g}(z)},
\end{equation}
as well as the large-scale galaxy bias:
\begin{equation}
    b_\mathrm{g}(z)=
    \int dM_\mathrm{h}\,
    b_\mathrm{h}(M_\mathrm{h})\,
    n(M_\mathrm{h},z)
    \frac{\langle N_\mathrm{g}(M_\mathrm{h})\rangle}{n_\mathrm{g}(z)}.
\end{equation}
and the satellite fraction:
\begin{equation}
    f_\mathrm{s} = 1 - \int dM_\mathrm{h} n(M_\mathrm{h},z) \frac{\langle N_\mathrm{c}(M_\mathrm{h}) \rangle}{n_\mathrm{g}(z)}
\end{equation}
Here, $b_\mathrm{h}(M_\mathrm{h})$ is the halo bias relation from \cite{Tinker2010}, and $n(M_\mathrm{h},z)$ is the halo mass function described by \cite{Tinker2008}. 
Satellite fraction:

The model galaxy number density is defined as
\begin{equation}
    n_\mathrm{g}(z)=
    \int dM_\mathrm{h}\,n(M_\mathrm{h},z)\langle N_\mathrm{g}(M_\mathrm{h})\rangle.
\end{equation}

Further details on the HOD formalism and its application to galaxy clustering analyses can be found in a comprehensive pedagogical overview presented by \cite{Asgari2023} and references therein.


\section{Results} \label{sec:results}

\begin{figure*}
    \centering
    \includegraphics[width=\linewidth]{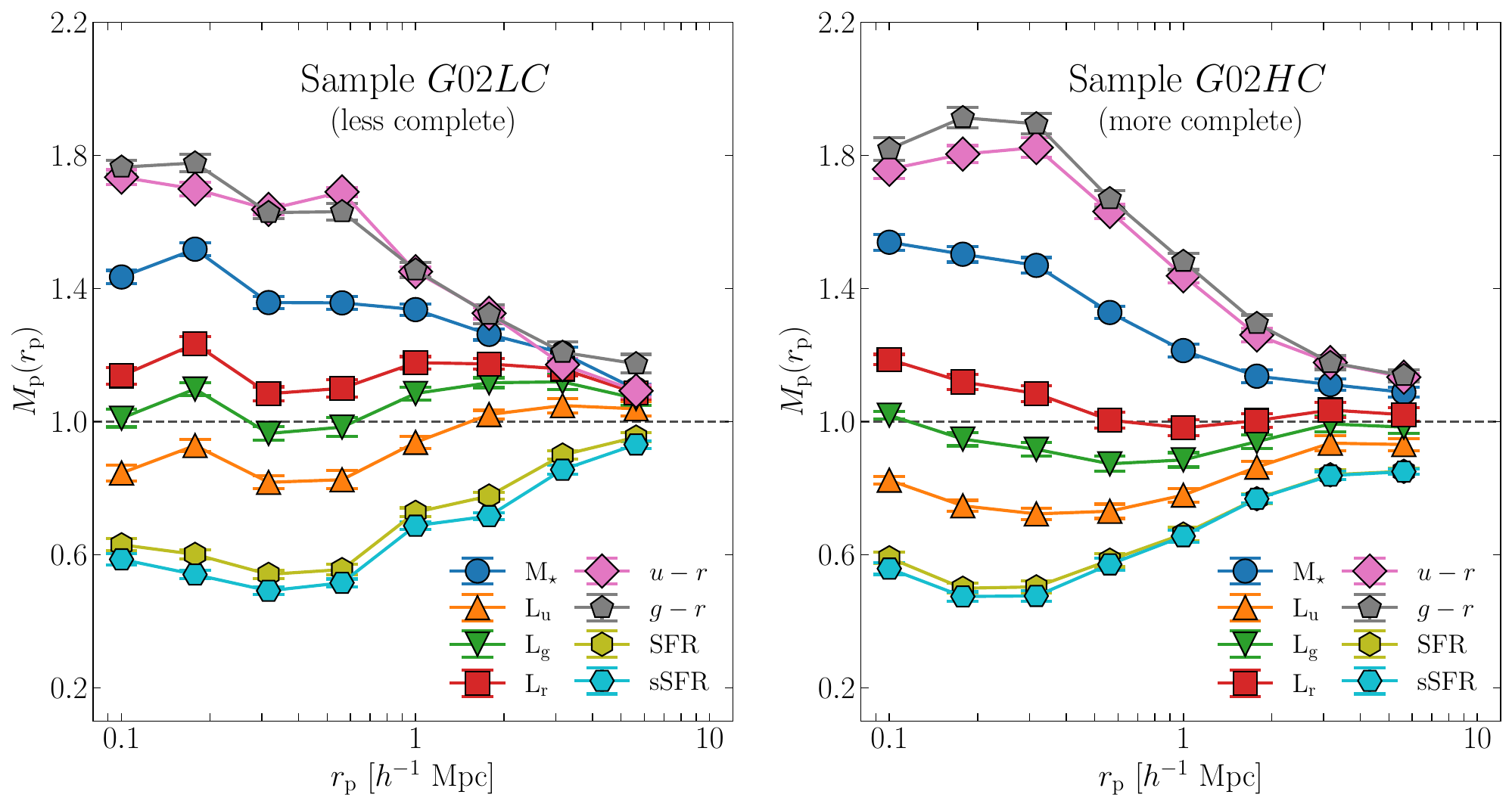}
    \caption{Rank-ordered projected MCFs for the low-completeness sample $G02LC$ (\textit{left panel}) and the high-completeness sample $G02HC$ (\textit{right panel}) within the G02 region of the GAMA survey.
    Different markers denote the different galaxy properties used in the MCF calculation, as indicated in the legend.
    The horizontal dashed line represents $\Mprp=1$.
    The MCF error bars shown are the jackknife and shuffling errors added in quadrature.
    }
    \label{fig:Region_G02_MCF}
\end{figure*}

\begin{figure*}
    \centering
    \includegraphics[width=\linewidth]{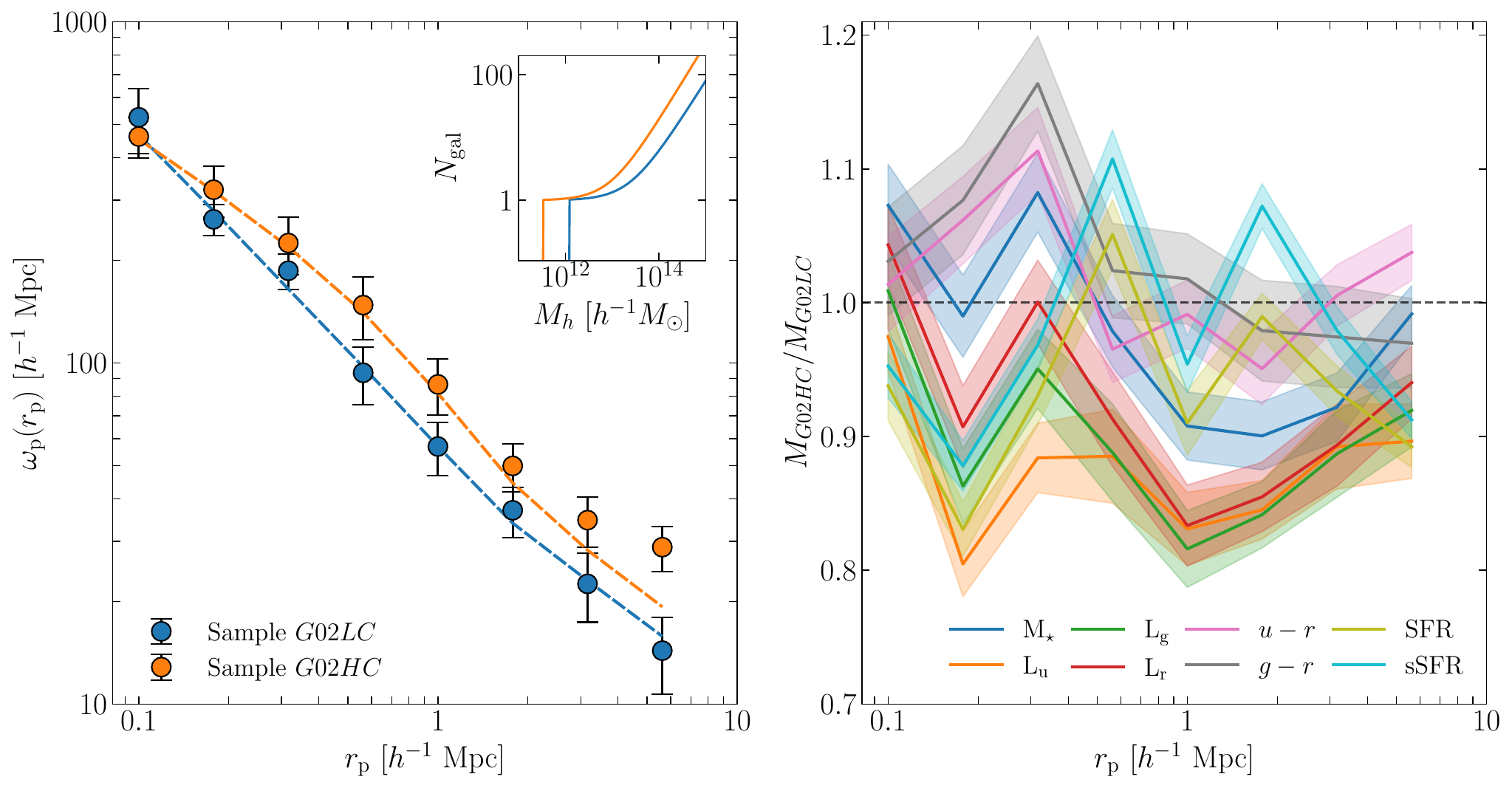}
    \caption{
    \textit{Left panel}: Projected 2pCFs for the $G02LC$ and $G02HC$ samples, with the HOD model fits. 
    The inset plot shows the corresponding halo occupation functions for both the samples, color-coded to match the main curves.
    \textit{Right panel}: Ratio of MCFs of the high-completeness sample to that of the low-completeness sample, $M_{G02HC}/M_{G02LC}$. 
    Different colored lines represent the MCF ratios for different galaxy properties.
    Error bars of $\wprp$ are estimated via jackknife resampling.
    While the errors in MCF ratios, calculated in quadrature, are represented by their corresponding color shaded region.
    }
    \label{fig:Region_G02_2pCF}
\end{figure*} 

\begin{figure*}
    \centering
    \includegraphics[width=\linewidth]{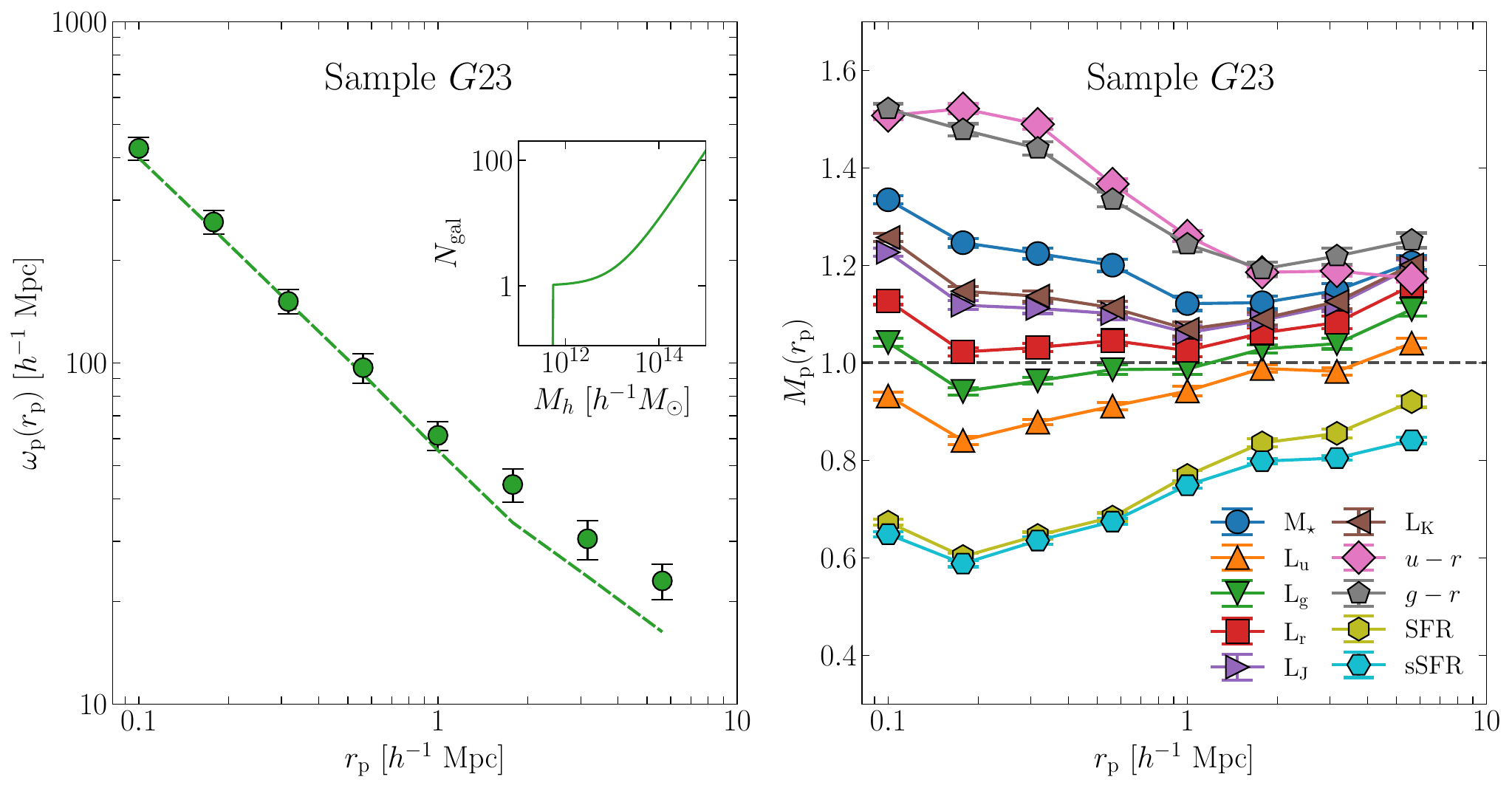}
    \caption{\textit{Left panel}: Projected 2pCF for sample $G23$, with the HOD model fits.
    The inset shows the corresponding halo occupation functions for the sample. 
    Error bars of 2pCF are from the diagonal elements of the covariance matrix obtained through jackknife resampling. 
    \textit{Right panel}: Rank-ordered projected MCFs with different markers indicating the different galaxy properties used as marks. 
    The MCF error bars shown are the jackknife and shuffling errors added in quadrature.
        The horizontal dashed line represents $\Mprp=1$.}
    \label{fig:Region_G23_2pCF_MCF}
\end{figure*}

In this section, we present the results of the 2pCF and MCF measurements in different samples of galaxies from the southern regions of the GAMA survey in the redshift range $0.1 < z < 0.2$.
All the correlation functions were measured for the projected separation scales of $0.1 < \rp/\hMpc < 7.5$.
For all the samples, MCFs were measured using the $u-r$ and $g-r$ colors, luminosities in $u, g, r$ bands, stellar mass, SFR, and sSFR as marks. 
Additionally for the sample $G23$, we used $J$-band and $K$-band luminosities as marks. 
The 2pCFs were fitted using power-law models and HOD model as described in Section~\ref{subsec:meth_fitting} and ~\ref{subsec:meth_HOD}. 
As described in Section~\ref{subsec:meth_err}, we used the jackknife method to estimate the uncertainties in 2pCFs and a combination of the jackknife method and random mark shuffling method in the case of MCFs.

\subsection{Correlation functions in the G02 region}\label{sec:results:g02}

In Fig.~\ref{fig:Region_G02_MCF}, we show the rank-ordered projected MCFs for low-complete sample $G02LC$ and high-complete sample $G02HC$.
These samples are selected from the G02 region of the GAMA survey and are volume-limited with a stellar mass limit $\mass~>~9.5$.
The first observation is that all the presented MCFs show a significant deviation from unity across all $r_\mathrm{p}$ scales.
This deviation is more pronounced at smaller scales and weakens with increasing scale.
A greater deviation of the MCF from unity indicates a stronger correlation between the corresponding property and the local environment.
This suggests that galaxy colors, luminosities, stellar mass, and SFR are strongly correlated with the local environment.
Different MCFs exhibit varying degrees of deviation from unity.
This indicates that the corresponding galaxy properties have different strengths in their correlation with the environment.

Among the properties considered, the color MCFs exhibit the greatest deviation from unity than those of other properties.
This indicates the strongest correlation between color and environment compared to all other properties considered.
It is also noted that the color MCFs show the highest amplitude (positive deviation from unity) among all the properties.
A positive deviation of the MCF from unity ($\Mprp~>~1$) indicates a stronger correlation between the corresponding property and the local environment, while a negative deviation ($\Mprp~<~1$) suggests an anti-correlation.
This suggests that the correlation between color and environment is positive, meaning that closer pairs of galaxies, on average, exhibit greater values of $u-r$ and $g-r$ colors.

The stellar mass MCF follows the colors in terms of its amplitude, making it the second most correlated property in this study.
Luminosity comes next, showing a hierarchy where the redder bands (e.g., $g$-band here) show a correlation with the environment, while bluer bands (e.g., $u$-band) show an anti-correlation.
The SFR and sSFR exhibit an anti-correlation across all the scales.
We provide an interpretation of these observations in Section~\ref{sec:discussions:envt}.

In Fig.~\ref{fig:Region_G02_2pCF}, we verify the influence of redshift completeness on clustering measurements by comparing the 2pCFs and MCFs between the samples $G02LC$ and $G02HC$.
We recall that these samples differ in their redshift completeness, with $G02LC$ having 31.6\% and $G02HC$ having 95.5\%.
The left panel shows the projected 2pCFs of the two samples.
The 2pCF measurements are depicted by the circles with errorbars and the dashed lines represent the HOD fits.
The inset plots show the corresponding halo occupation functions.

The power-law fitting gives correlations lengths of $5.88\pm0.67 \ \hMpc$ and $4.90\pm0.42 \ \hMpc$ and slopes of $1.67\pm0.05$ and $1.91\pm0.06$ for $G02HC$ and $G02LC$ samples, respectively.
We notice that the $r_0$ parameters of the high-completeness and low-completeness samples differ by $1.4\sigma$, and the $\gamma$ parameters differ by $3.1\sigma$. 

We further investigate these differences by the HOD modeling of the correlation functions for samples $G02LC$ and $G02HC$ (shown in left panel of Fig. \ref{fig:Region_G02_2pCF} and Table \ref{tab:HODparams}). The incompleteness systematically affects the best-fit HOD parameters, shifting the characteristic mass scales $M_{\mathrm{min}}$ and $M_1$. The obtained characteristic masses of $\log M_{\mathrm{min}}$ and $\log M_{1}$ differ by 0.55, 0.43 dex, respectively, corresponding to differences of more than $5\sigma$. However, this systematic shift does not significantly affect the effective halo masses or satellite fractions which remain consistent between the two samples. Hence, incompleteness biases the individual HOD parameters but not necessarily the integrated halo properties of the sample. We further discuss these results in Section \ref{sec:discussions:zcompleteness}.

In the case of MCFs, we observe that they follow the same trend in both the samples for all the properties. 
However, the MCF amplitudes between the two samples vary significantly.
The right panel of Fig.~\ref{fig:Region_G02_2pCF} shows the ratios of MCFs of the high-completeness sample $M_{G02HC}$ to that of the low-completeness sample $M_{G02LC}$.
In general, we observe a systematic decrease in the MCF amplitudes in the high-completeness sample relative to the low-completeness sample.
This effect of varying redshift completeness is further discussed in Section~\ref{sec:discussions:zcompleteness}.

\subsection{Correlation functions in the G23 region}\label{sec:results:g23}

In Fig.~\ref{fig:Region_G23_2pCF_MCF}, we show the projected 2pCF and rank-ordered MCF for sample $G23$.
This sample has been selected from the G23 region of the GAMA survey and is volume-limited with a stellar mass limit $\mass > 9.7$.
In the left panel of Fig.~\ref{fig:Region_G23_2pCF_MCF}, the circles represent the 2pCF measurements, and the dashed line indicates the HOD fit to these measurements.
The inset plots show the corresponding halo occupation functions.
In power-law fitting, we obtain a correlation length of $5.05\pm0.52 \ \hMpc$ and a slope of $1.75\pm0.07$.


In terms of halo occupation, see left panel of Fig. \ref{fig:Region_G23_2pCF_MCF} and Table \ref{tab:HODparams}, we find the effective halo mass of G23 sample to be $\log (M_{\mathrm{eff}} / h^{-1} \mathrm{M_{\odot}}) = 13.33\pm0.02$. This value is consistent with the average halo masses estimates obtained from weak gravitational lensing studies of GAMA equatorial fields by \cite{Viola2015}. We also obtain the large scale galaxy bias $b_g = 1.13\pm0.01$.


The right panel of Fig.~\ref{fig:Region_G23_2pCF_MCF} shows the MCFs obtained for the G23 galaxies using $u-r$ and $g-r$ colors, stellar mass, luminosities in $u, g, r, J, K$ bands, SFR, and sSFR as marks.
Similar trends are observed in these MCFs as in the G02 regions (Fig.~\ref{fig:Region_G02_MCF}).
Among all the MCFs, the $u-r$ and $g-r$ color MCFs exhibit the highest deviation from unity, followed by the stellar mass MCFs.
In the G23 region, we include luminosities in the $K$ and $J$ bands in addition to those presented for the G02 region.
We observe a similar hierarchy among the optical bands, where the environmental correlation transitions into an anti-correlation from redder to bluer bands. 
The SFR and sSFR MCFs show an anti-correlation across all scales.

We compare the measurements in the GAMA southern regions to those in the equatorial regions in Fig.~\ref{fig:A1_comparison} and discuss in Section~\ref{sec:discussions:comp_equatorial}.
Unlike in the samples from G02 and equatorial regions \citep{Unni_GAMA}, we observe a slight rise in MCFs at $\rp \sim 7\ \hMpc$ in the G23 region.
A possible explanation for this observation is discussed in Section~\ref{sec:discussions:comp_equatorial}.

\section{Discussions} \label{sec:discussion}

\subsection{Environmental dependence of galaxy properties}\label{sec:discussions:envt}

Galaxy color is an effective tracer of galaxy environment \citep{Blanton2005,Zehavi2011,Skibba2014}.
It is known to be a reliable indicator of cluster- and group-centric distance, with the fraction of bluer galaxies increasing as the distance from the group center grows \citep{MartinezMuriel2006}. 
This is expected, as environmental quenching through processes like ram-pressure stripping \citep{GunnGott1972} and tidal disruptions \citep{Merritt1983} is more pronounced at the centers of clusters and groups, where gas density and temperature are higher. 
In studies where the influence of environment on morphology has also been looked at, it is found that though galaxies become redder with increasing density, their morphology remains largely unchanged \citep{Ball2008}, implying that it is the environment, rather than morphology, that is primarily responsible for the quenching of star formation in galaxies and the subsequent reddening of their colors \citep{Sotillo-Ramos2021}. 
The MCF behaviors observed for our sample conform to these trends. The $u-r$ and $g-r$ colors yield the strongest environmental signal, with redder galaxies exhibiting enhanced clustering relative to bluer populations.
A discussion on the environmental dependence of color in the context of halo occupation is given in Sec.~\ref{sec:discussion:color}.

In addition, galaxies in overdense environments tend to possess large stellar mass due to enhanced rates of dry mergers and interactions with other massive cluster members, leading to their formation \citep[e.g.,][]{kauffmann2004}. The stellar populations of such galaxies also tend to be older due to the accelerated quenching of star formation through environmental processes \citep{peng2010}. Such a trend is also evident in the MCF of stellar mass for galaxies in our sample, with higher stellar mass systems more strongly clustered than their lower-mass counterparts.

In contrast, the MCF trends for SFR and sSFR show an inverse relationship with galaxy environment. 
The $u$-band luminosity, which traces young stellar populations (and therefore can be an SFR indicator), also follows a similar trend, as also noted in previous studies \citep{Deng2012,Unni_GAMA}. 
These trends are consistent with observational studies reporting a decline in the median sSFR of star-forming galaxies with increasing environmental overdensity \citep[e.g.,][]{Patel2009,Patel2011}.

The combined MCF trends of stellar mass and sSFR suggest that galaxies residing in overdense regions of the large-scale structure tend to be massive with suppressed star formation.
This trend is consistently observed in galaxy surveys, with low-density regions dominated by low-mass, high-sSFR galaxies, and high-density regions hosting a substantial population of massive galaxies with old stellar populations \citep{kauffmann2004,McGee2011,Muzzin2012,Wang2018}. 
As simulations show, this transition in sSFR with density is a natural fallout of galaxies losing their gas due to the cluster environment. 
In the TNG50 simulations, \cite{Ding2024} find that the rate of such environmentally induced quenching can be rather high, with as many as 42\% of galaxies in a cluster losing as much as 80\% of their cold gas within a billion years of infall into the cluster.

The MCF results also show that galaxies brighter in the $g$, $J$, and $K$ bands are more densely grouped compared to galaxies that are bright in the $u$-band, with the $K$-band luminosity tracing the environment the best. 
The $K$-band luminosity MCF displays trends analogous to those of the stellar mass MCF, albeit with a reduced amplitude. 
This indicates a strong correlation between $K$-band luminosity, stellar mass, and environmental density. 
Therefore, the close similarity between the stellar mass and $K$-band luminosity MCFs suggests that, in the absence of stellar mass information for comparison, $K$-band luminosity can serve as a reliable proxy for stellar mass in galaxy clustering studies \citep{Baldry2012}.
This has important implications for current and next-generation NIR-selected surveys.
For JWST deep fields, and wide-area missions such as Euclid and the Roman Space Telescope, $K$-band (or related NIR) photometry would offer a first-order estimate of stellar mass distributions without immediate spectroscopic follow-up, thereby greatly enhancing survey efficiency.


\subsection{Environmental dependence of galaxy color}\label{sec:discussion:color}
To study in more details the color-environment dependence, we split sample $G23$ into red and blue galaxies according to their $g-r$ color, with the dividing threshold equal to 0.83 (being the median of $g-r$ distribution for all galaxies). This division creates two distinguishing galaxy populations: blue star-forming and less massive sample with median mass $\log (M_{\star}/\mathrm{M_{\odot}}) = 10.17$ and $\log {\rm sSFR} = -9.86$, and red quiescent more massive sample with median mass $\log (M_{\star}/\mathrm{M_{\odot}}) = 10.70$ and  $\log {\rm sSFR} = -11.3$ (see Table \ref{tab:prop_samples}).

In Figure \ref{fig:HODcolor} we present the 2pCF measurements for the red and blue samples, along with the best-fit HOD model and the corresponding halo occupation functions. As shown, red galaxies cluster more strongly than blue galaxies. This is especially evident on small scales of $\lesssim 1-2$  $h^{-1} \mathrm{Mpc}$, i.e., within individual halos. This is likely because red galaxies preferably occupy rich clusters, resulting in larger number of central-satellite and satellite-satellite pairs. Indeed the HOD modeling of the galaxy corelation functions measured for both samples show a clear color dependence of galaxy-halo connection. Red galaxies are found to be both centrals and satellites in more massive halos, which is reflected in their higher $M_\mathrm{min}$ and $M_1$ (see Table \ref{tab:HODparams}). The higher effective mass and galaxy bias, also both indicate that red galaxies trace dense regions of the cosmic web. These results are in agreement with the literature \citep[e.g.,][]{Li2006, Coil2008, Zehavi2011, Guo2013, Paul2019}.


Moreover, we obtain a significantly higher satellite fraction for red galaxies compared to blue galaxies, with $13\%$ and $7\%$ respectively, which is consistent with previous studies \citep[e.g][]{Zehavi2011, Xu2018}. These results, together with the measurements of the higher effective halo masses for red sample, imply that red galaxies are more frequently satellites in massive halos, while blue galaxies are predominantly central galaxies in less massive halos. 

To complete this picture, we can take a look at the characteristic halo masses ratio, $M_1/M_\mathrm{min}$, which gives us information how halo occupation depends on galaxy color. We found that the halo hosting, on average, one red central galaxy and one red satellite red galaxy needs to be 24 times more massive than the halo that hosts only one central red galaxy. For blue galaxies, this ratio is almost three times higher and reaches 64. The higher $M_1/M_\mathrm{min}$ ratio for blue galaxies means that a halo needs to grow considerably more massive before it hosts, on average, one blue satellite galaxy. This implies two things. First, it confirms that blue galaxies are predominantly centrals in isolated halos located in low-density environments where halo mergers are rare \citep[e.g][]{Fakhouri2009}, allowing galaxies to evolve without the quenching influence of the environment. Second, the relatively small halo mass gap for red galaxies means that halos hosting red galaxies begin to accumulate satellites at much lower masses. This is a natural consequence of their preference for high-density environment, where halo mergers are frequent. After a halo merger, galaxies that become satellites are relatively quickly quenched by ram-pressure stripping, tidal interactions, and/or galaxy harassment \citep{GunnGott1972,Merritt1983}. This is also why we observe such a strong correlation of {\it g-r} color and the environment in the marked correlation function.

\begin{figure}
    \centering
    \includegraphics[width = 0.45\textwidth]{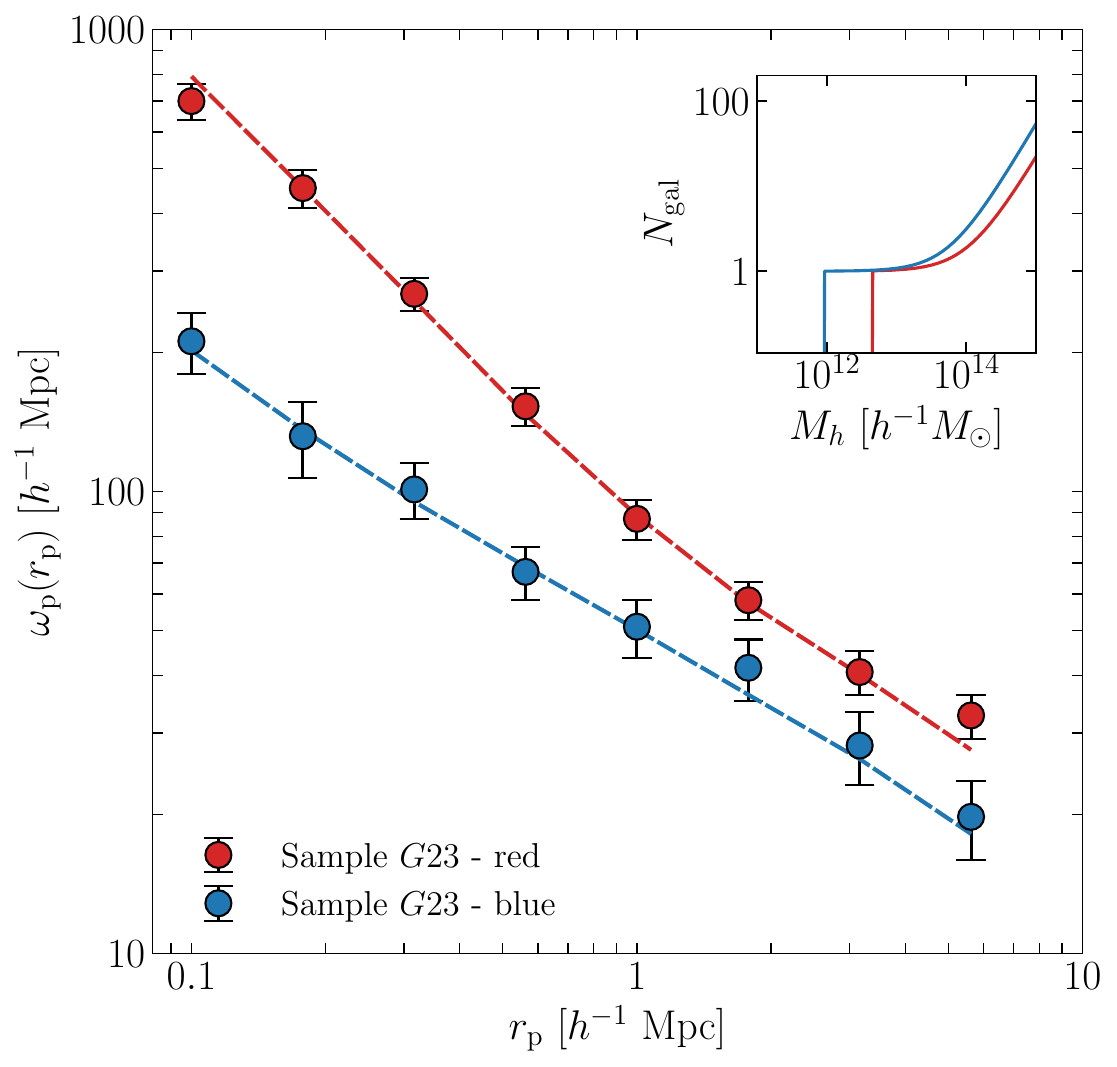}
    \caption{Projected 2pCF for color selected samples from G23 field, with the HOD model fits. The inset show the corresponding halo occupation functions for each color-selected sample.}
    \label{fig:HODcolor}
\end{figure}


\subsection{Redshift incompleteness effect}\label{sec:discussions:zcompleteness}

Redshift incompleteness could significantly influence the galaxy clustering measurements \citep{meng2024}.
In this work, the samples $G02LC$ and $G02HC$ differ in their redshift completeness, providing an opportunity to assess the impact of incompleteness on our results.
In Section~\ref{sec:results:g02}, we observed that the 2pCF and MCF measurements are significantly affected due to the difference in the redshift completeness.
We note a systematic fall in the MCF amplitudes for the high-completeness sample compared to the low-completeness sample. 
This suggests that the low-complete sample shows a stronger environmental dependence than the high-complete sample.
These results highlight the importance of accounting for redshift incompleteness effects, which can possibly influence the clustering measurements.




To look for the physical origin of the clustering differences between the complete and incomplete samples we interpret the measurements within the framework of the galaxy-halo modeling. As seen in Fig.~\ref{fig:Region_G02_2pCF}, the correlation functions measured for both samples are nearly identical at small scales $r_\mathrm{p} \sim 0.1 - 0.3 h^{-1} \mathrm{Mpc}$, which means that clustering within individual halos, remains the same for both samples. On the other hand, at the intermediate and large scales $r_\mathrm{p} > 0.5$ $h^{-1} \mathrm{Mpc}$ the correlation functions show substantial differences (of on average $\sim30\%$), with incomplete sample having lower values of $\omega_p(r_p)$. The incomplete sample also has a higher galaxy bias, $b_\mathrm{g} = 1.25\pm0.01$, compared to $b_\mathrm{g} = 1.15\pm0.03$ for the complete sample. This suggests that the incomplete sample traces stronger over-densities of the large scale structure.

To explain these observations, we first need to note that the incompleteness in sample $G02LC$ is not random, but is mostly driven by the shorter observational exposure. As a result, this sample lacks low stellar mass (usually fainter) galaxies that predominantly occupy less massive halos. As shown in Fig. \ref{fig:Region_G02_mass_distribution} the stellar mass distribution of the G02LC sample is shifted toward higher masses, with a median stellar mass of $\log (M_{\star}/{\rm M}_{\odot}) = 10.41$ compared to $\log (M_{\star}/{\rm M}_{\odot}) = 10.29$ for the complete sample. This shift is particularly evident at the low stellar mass end, where the first quartile is $\log (M_{\star}/{\rm M}_{\odot}) = 9.87$ compared to $\log (M_{\star}/{\rm M}_{\odot}) = 9.82$ for the complete sample.

\begin{figure}
    \includegraphics[]{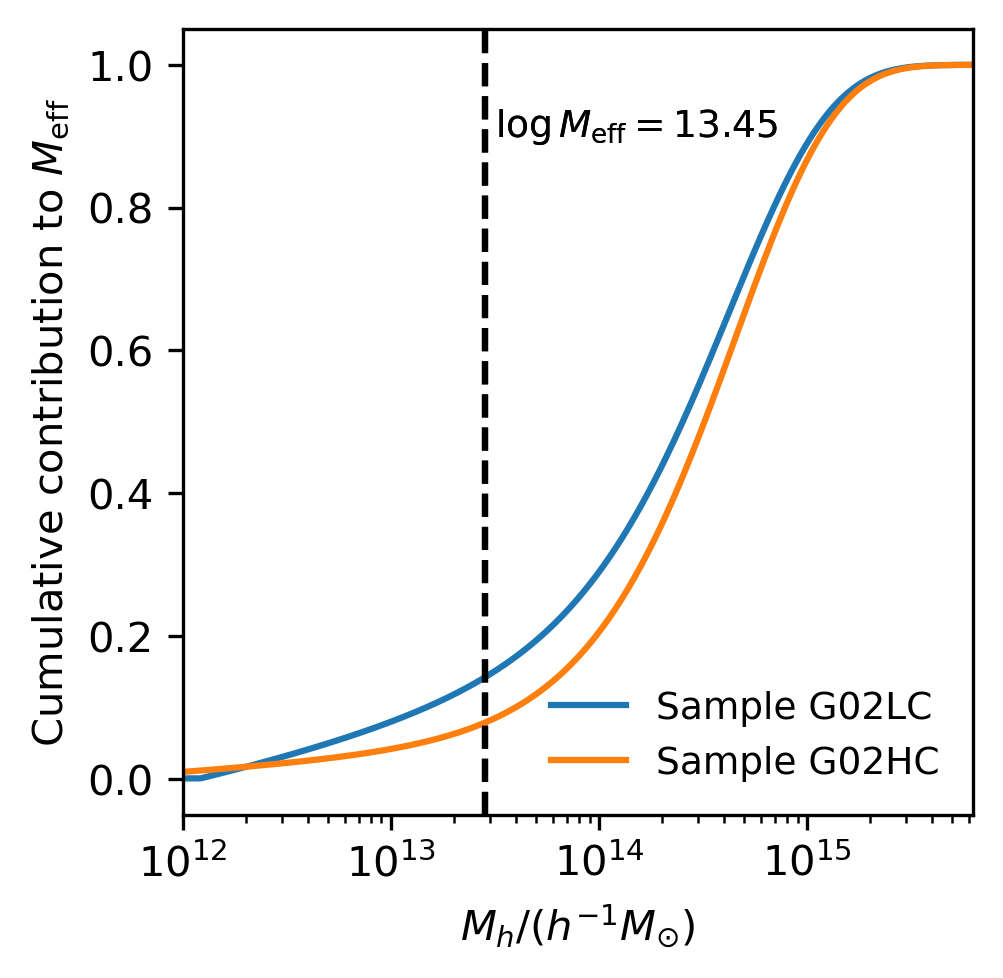}
    \caption{Cumulative contribution to $M_{\mathrm{eff}}$ as a function of the halo mass for complete (G02HC, orange) and incomplete (G02LC, blue) samples. Black, dashed line mark the effective halo mass obtained from the best-fit HOD models, which is the same for both samples.}
    \label{fig:HODcumulative}
\end{figure}

The lack of low mass galaxies in the incomplete sample affects both the shape of the correlation function and the HOD parameters estimates. The most evident effect is the reduced amplitude of the correlation function on the intermediate and large scales due to the missing pairs of less massive galaxies. In the framework of the HOD, the lack of low-mass galaxies results in higher estimates of characteristic halo masses $M_\mathrm{min}$ and $M_1$. It is well established that more massive galaxies are expected to reside in more massive halos \cite[see e.g,][]{Zehavi2011, Leauthaud2012, Behroozi2019}. Unsurprisingly then for the incomplete sample, consisting predominantly of more massive galaxies, $M_\mathrm{min}$ and $M_1$ are systematically higher by $\sim0.5$ dex than those for the complete sample (see Table \ref{tab:HODparams}).

However, despite these differences, the estimates of the effective halo mass $M_{\mathrm{eff}}$ for both, the complete and incomplete, samples are identical, with $\log (M_{\mathrm{eff}}/ h^{-1} {\rm M_{\odot}}) = 13.45$. 
This indicates that the sample incompleteness mainly changes the halo abundance, rather than the effective host halo mass. The galaxies that remain in the incomplete sample still inhabit essentially the same halo population. 

The primary reason is that the $M_{\mathrm{eff}}$ is an integral quantity, making is less sensitive (compared to $M_1$ and $M_\mathrm{min}$) to changes of sample stellar mass threshold. The incompleteness of sample $G02LC$ mainly concerns low mass galaxies inhabiting lower-mass halos, whose contribution to the effective halo mass is small. As presented in Fig. \ref{fig:HODcumulative} the halos with masses below $< 10^{13}$ contribute only $\sim 10\%$ to the cumulative effective halo mass. The unchanged estimate of $M_{\mathrm{eff}}$ therefore suggests that the galaxies missing from the incomplete sample are most likely central galaxies in low-mass halos, which is also supported by the unchanged satellite fraction. Removing them from the sample increases $M_\mathrm{min}$, but because they contribute relatively little to $M_\mathrm{eff}$ the overall halo population remains unchanged.

\subsection{Stellar mass incompleteness effects}\label{sec:discussions:masscompleteness}

\begin{figure}
    \includegraphics[width = 0.45\textwidth]{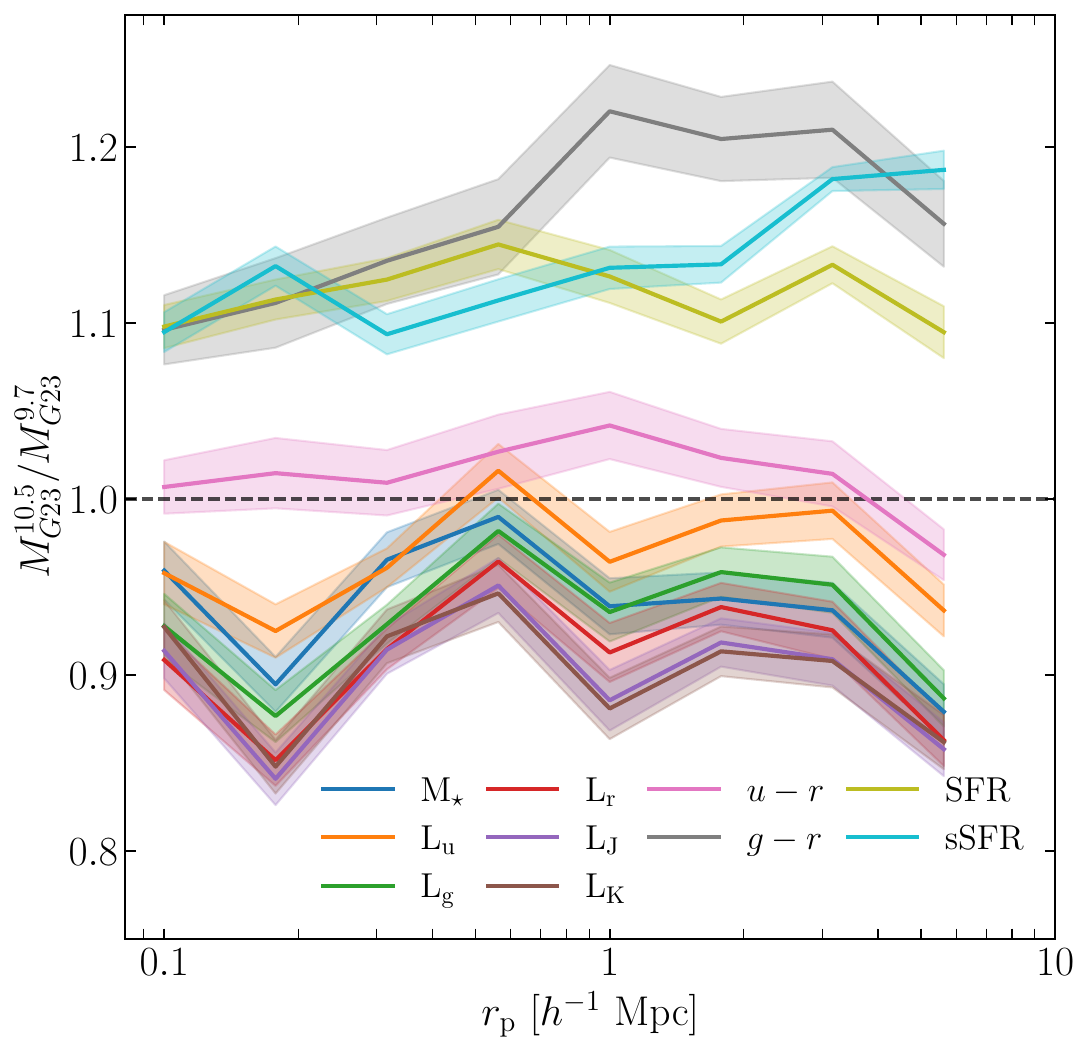}
    \caption{Projected MCFs comparison between two samples of sample $G23$ with more stellar mass completeness (stellar mass cut of 10.5) and less stellar mass incompleteness (stellar mass cut of 9.7). Different galaxy property MCFs are represented by different colored lines and errors in MCF ratios is represented by corresponding color shaded region.}
    \label{fig:mass_incompleteness}
\end{figure}

In this section, we investigate the effect of stellar mass incompleteness on the results. 
All the samples used in this work are stellar mass selected flux-limited samples, and are hence inevitably stellar mass incomplete. 
To investigate the effects of stellar mass incompleteness, we remeasure MCFs in the sample $G23$ with an increased stellar mass limit of $\log (M_{\star}/{\rm M}_{\odot}) = 10.5$.
We present the MCF ratio between these samples in Fig  \ref{fig:mass_incompleteness}.
We see that all the galaxy properties except $u-r$, $g-r$, SFR, and sSFR have lower amplitudes, and consequently less correlation, compared to the incomplete sample.
Although the MCF ratios for SFR and sSFR are above unity (indicating higher amplitudes), this actually reflects a decrease in their anti-correlation relative to the incomplete sample.
This behavior mirrors our findings from the redshift completeness comparison -- the incomplete sample demonstrates higher correlation amplitudes for all galaxy properties except for the $u-r$ and $g-r$ colors.
This aligns with the findings of \citet{Unni_GAMA}, who noted that a stellar mass-incomplete sample exhibited stronger correlations than a more complete counterpart.

However, while \citet{Unni_GAMA} did not explore color MCFs, our analysis includes them and reveals that color correlations strengthen in a more stellar mass-complete sample, opposite to the behavior observed for other galaxy properties.
This phenomenon arises because in a flux-limited survey, the stellar mass completeness limit depends on color and redder galaxies are missed in stellar mass-incomplete samples \citep{taylor2011_gama_stellar_mass}.
The redder galaxies have systematically higher stellar mass-to-light ratios than bluer galaxies \citep{Bell2001}.
This is particularly relevant because \citet{Du2020} demonstrated that the stellar mass-to-light ratio correlates positively with both $g-r$ and $u-r$ colors.
Notably, the $g-r$ color MCF displays a larger increase in amplitude than $u-r$; a similar trend is reported by \citet{Du2020}, where stellar mass-to-light ratio for $g$ and $r$-bands show stronger correlations with $g-r$ than with $u-r$.

As a result, a sample with higher stellar mass-completeness exhibits a higher mean stellar mass-to-light ratio, and hence include more redder galaxies compared to its incomplete counterpart.
This test tells us that if our samples were more stellar mass complete then we would see a lesser correlation for all luminosities, mass, SFR, and sSFR as compared to the current analysis, whereas for $u-r$ and $g-r$ the correlation would increase further.

\subsection{Comparison with GAMA equatorial regions}\label{sec:discussions:comp_equatorial}

\begin{figure*}
    \centering
    \includegraphics[width=\linewidth]{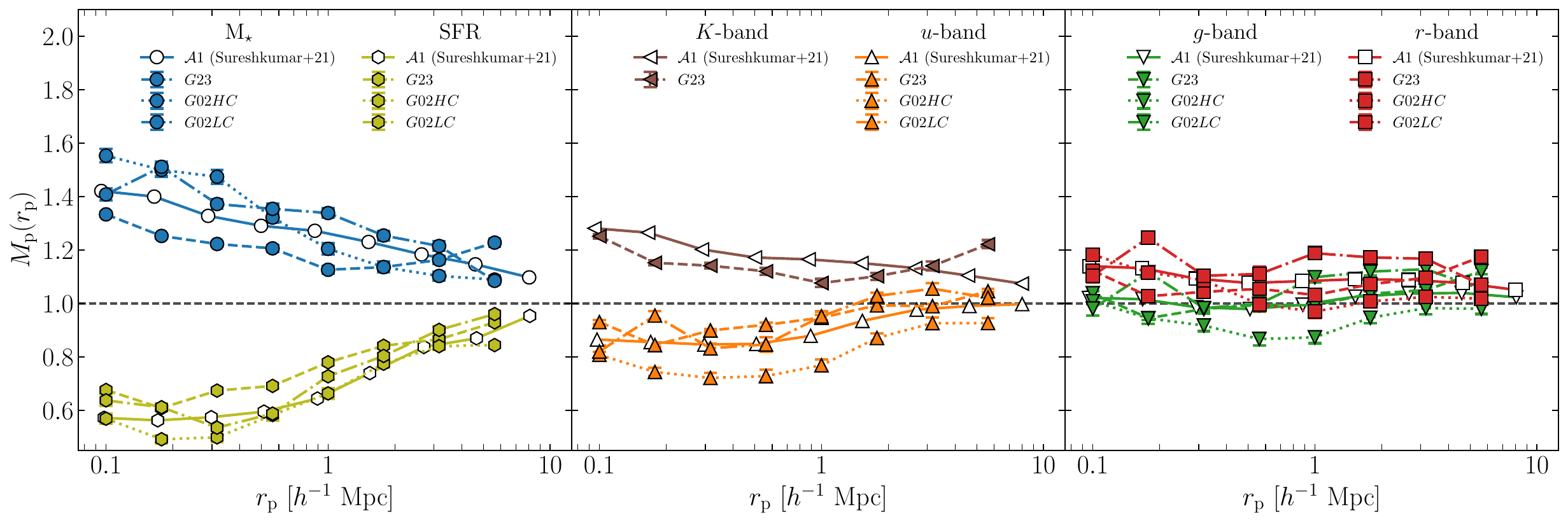}x
    \caption{Comparison of the projected MCFs in the southern regions (sample $G23$: dashed line, sample $G02HC$: dotted line, and sample $G02LC$: dashed dotted line) with those in the equatorial regions (sample $\mathcal{A}1$: solid line). The MCFs for the equatorial regions are taken from sample $\mathcal{A}1$ of \cite{Unni_GAMA}. Each panel shows the comparison for two galaxy properties (\textit{left}: stellar mass and SFR, \textit{center}: $K$-band and $u$-band, and \textit{right}: $g$-band and $r$-band). Note that the error bars for sample $\mathcal{A}1$ are too small to be visible. 
    }
    \label{fig:A1_comparison}
\end{figure*}


In this section, we compare the MCFs obtained for the southern GAMA regions in this work with those from the equatorial GAMA regions by \citet{Unni_GAMA}.
In Fig.~\ref{fig:A1_comparison}, we compare the MCFs of various galaxy properties measured in our samples $G02LC$, $G02HC$, and $G23$ with those from sample~$\mathcal{A}1$ of \citet{Unni_GAMA}.
The left panel shows MCFs for stellar mass and SFR, the center panel displays MCFs for the $K$- and $u$-bands, and the right panel includes the $g$- and $r$-band results.
We do not include comparisons for the $J$-band and sSFR MCFs, as their amplitudes and trends closely resemble those of the $K$-band and SFR, respectively.
Additionally, since \citet{Unni_GAMA} did not compute MCFs for colors, we omit comparisons of color MCFs.

We find that the qualitative trends observed in the southern regions -- the relative hierarchy of properties in their correlation with environment -- are consistent with those in the equatorial fields.
The MCF amplitudes, however, differ significantly between equatorial and southern samples given the errorbars. 
These differences are likely driven by differences in redshift and stellar mass cuts, redshift completeness, random catalogue generation techniques, and cosmic variance due to local structures.
Additionally, we notice slight deviation in the G23 field at larger scales, which is discussed below.




In Fig.~\ref{fig:Region_G23_2pCF_MCF}, we observe that the MCF measurements in the sample $G23$ show an elevated amplitude on separations $r_\mathrm{p} \sim 7\ \hMpc$, unlike in the G02 samples.
This suggests that the correlation between galaxy properties and environment in the G23 region is enhanced at intermediate scales. 
A possible explanation for this behavior is the presence of local structures in the G23 region, as also noted by \citealt{Driver2022}. 
In the top right region of the RA-DEC plot for G23 (right panel of Fig.~\ref{fig:Region_G23_mass_distribution}), we note the presence of prominent local structures, which could influence the clustering signal.

A similar deviation was reported by \citealt{Driver2022} in the galaxy stellar mass function of G23 compared to other GAMA regions. 
They found that G23 exhibited both a higher overall mass density and a steeper low-mass upturn, similar to what is observed in the Virgo cluster. 
As argued by the authors, this could indicate that the G23 region intersects with a loosely bound structure that is yet to be fully characterized.
Furthermore, \citealt{Driver2022} found that applying an LSS correction to the stellar mass function reduces these deviations and brings G23 into closer agreement with the other GAMA fields. 
This suggests that the observed anomalies in both the stellar mass function and the MCF in the G23 region may arise from the same underlying LSS effects.




\section{Conclusion} \label{sec:conclusion}


The MCF analysis shows that different galaxy properties correlate with the local environment differently at separation scales of $r_\mathrm{p}<7.5\ \hMpc$.
The MCFs for all properties deviate from unity, with stronger deviations on small scales, highlighting the robustness of these correlations. Overdense regions of the large-scale structure are preferentially populated by massive, redder galaxies with lower star-formation rates.
Among the properties examined, rest-frame colors ($u-r$ and $g-r$) show the strongest deviation of the MCFs from unity, followed by stellar mass and red-band ($g$, $J$, and $K$) luminosities, while SFR and sSFR show consistent anti-correlations with the density of the environment.

The HOD modeling reveals a clear color dependence in the galaxy-halo connection. 
Red galaxies exhibit a higher satellite fraction (13\%) compared to blue galaxies (7\%) and require a substantially lower characteristic halo mass gap ($M_1/M_\mathrm{min}$) to accumulate satellites, reflecting their preference for high-density environments where halo mergers are frequent.

We further use HOD modeling to interpret the clustering strength differences between samples of differing redshift completeness.
The redshift incomplete sample under represents low stellar mass galaxies that occupy low mass halos, which reduces the amplitude of projected 2pCF at intermediate and large scales.
This also decreases the characteristic halo masses ($M_\mathrm{min}$ and $M_1$), although the effective halo mass and satellite fraction remain unchanged.

Both redshift and stellar mass incompleteness artificially alter clustering measurements. While a lack of completeness generally increases MCF amplitudes, indicating a stronger environmental correlation, color MCFs behave oppositely. Stellar mass complete samples display stronger color-environment correlations because they preserve galaxies with high stellar mass-to-light ratios.


This work presents the first clustering measurements of galaxies in the southern regions of the GAMA survey.
It builds on previous environmental studies \citep{Unni_GAMA,Unni2023} done on the equatorial regions of the GAMA survey, completing the full environmental analysis of the GAMA survey.
The similarity of the 2pCF values and MCF trends between the southern GAMA regions and the equatorial regions \citep{Unni_GAMA} indicates consistent agreement in our measurements of environmental correlations of galaxy properties across the GAMA regions, although the MCF amplitudes differ between the two regions.
We attribute this difference to differences in redshift ranges, stellar mass cuts, and redshift completeness levels, and cosmic variance due to the presence of local structures.
In region G23 specifically, we find that the local structures enhance the correlation between galaxy properties and environment at intermediate scales.

The observed dependence of galaxy color and the near-IR luminosities on environment can be used by near-IR surveys to trace the build-up of stellar mass and environmental quenching across redshifts.
Moreover, the amplitude and scale dependence of the measured correlations provide a critical benchmark for cosmological simulations trying to reproduce the dependence of galaxy assembly on both stellar mass and the large-scale environment. 
A direct comparison of these MCF measurements against mock catalogues from cosmological simulations such as IllustrisTNG (e.g.,\citealt{Shreeram2025}) would offer a stringent test of current galaxy formation models.
A complementary avenue is to construct a halo model that directly reproduces the observed MCF distributions themselves, extending existing luminosity- and color-based formalisms (e.g., \citealt{Skibba2006,Skibba2009}) to the full set of marks considered here: stellar mass, SFR, and near-IR luminosity.


\begin{acknowledgments}
We thank the anonymous referee for the useful comments and suggestions.
U.S. acknowledges support from the Polish National Science Center (NCN) grant UMO-2024/53/B/ST9/00230 and National Research Foundation of South Africa (grant no. 137975).
A.D. and A.P. are supported by the National Science Centre grant MAESTRO 2023/50/A/ST9/00579.
The authors thank Daniel Farrow for discussions on random samples, and Soumya Shreeram for discussion on TNG simulations.
GAMA is a joint European-Australasian project based around a spectroscopic campaign using the Anglo-Australian Telescope. The GAMA input catalogue is based on data taken from the Sloan Digital Sky Survey and the UKIRT Infrared Deep Sky Survey. Complementary imaging of the GAMA regions is being obtained by a number of independent survey programmes, including GALEX MIS, VST KiDS, VISTA VIKING, WISE, Herschel-ATLAS, GMRT and ASKAP, providing UV to radio coverage. GAMA is funded by the STFC (UK), the ARC (Australia), the AAO, and the participating institutions. The GAMA website is \url{https://www.gama-survey.org/}.
\end{acknowledgments}

%






\appendix

\section{Completeness Mask} \label{app:completeness_mask}

The G02 region has varying levels of redshift completeness, and this should be reflected in the random catalogues as well.
We account for this by generating a completeness map based on the method outlined in \citealt{Baldry2018}.
As shown in Fig~\ref{fig:Region_G02_completeness} we divide the region into $0.2\degr \times 0.2\degr$ cells, and each cell gets a value in the range $[0,1]$ based on the ratio of reliable redshift targets (\texttt{nQ} $\geq 3$) to SDSS selected targets (\texttt{SURVEY\_CLASS} $\geq 5$).
Using this completeness map, we create the random sample for $G02LC$, then imposing stellar mass cut of $\mass > 9.5$ that corresponds to the real sample.

\begin{figure}
    \centering
    \includegraphics[width=0.7\linewidth]{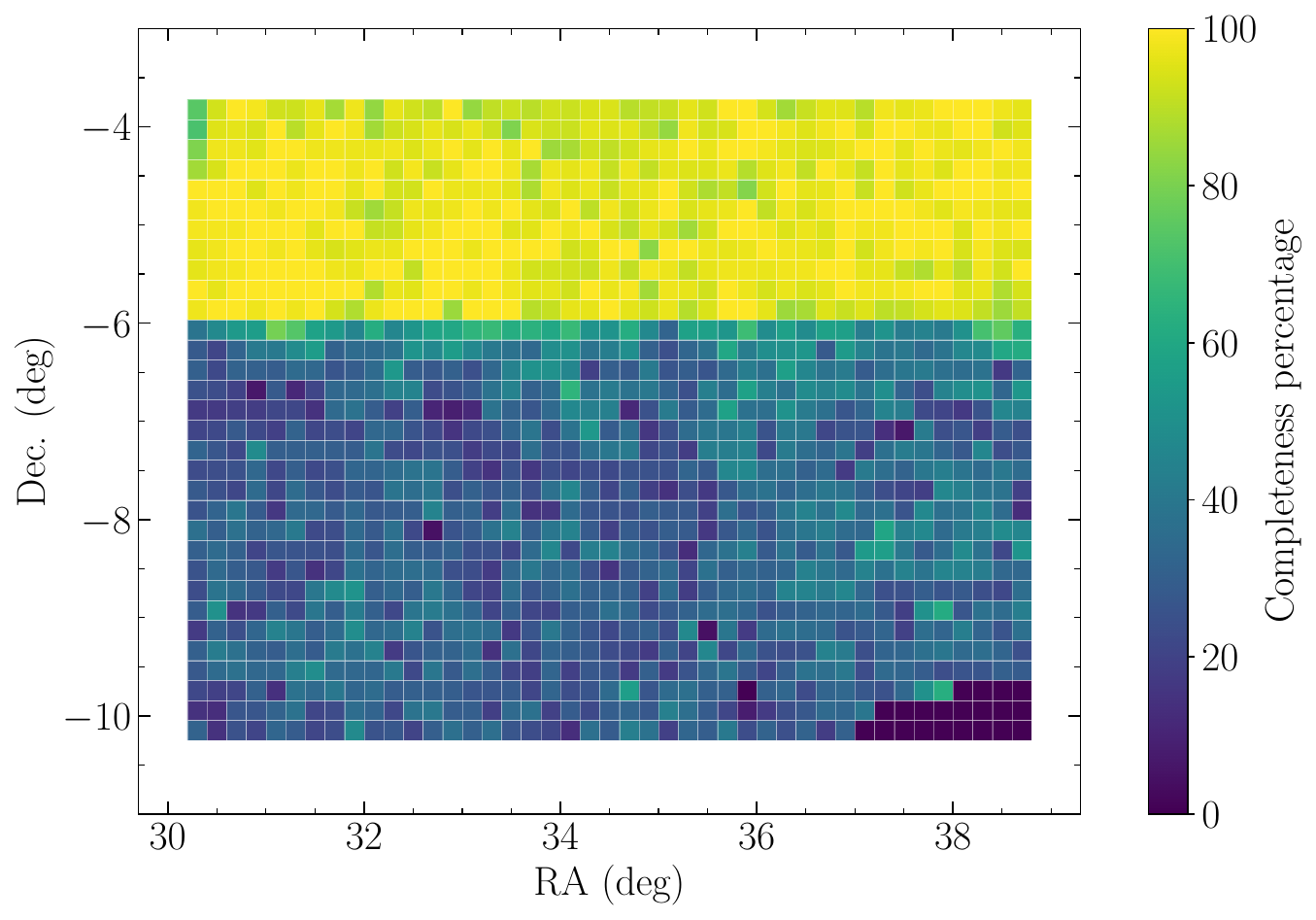}
    \caption{Redshift completeness map of region G02. Each cell covers an area of $0.2\degr \times 0.2\degr$, color coded by the ratio of reliable redshift targets (\texttt{nQ} $\geq 3$) to SDSS selected targets (\texttt{SURVEY\_CLASS} $\geq$5).}
    \label{fig:Region_G02_completeness}
\end{figure}

\section{Comparison of MCFs from \textsc{ProSpect} and \textsc{bagpipes} SFRs}\label{app:comparision_SFRs}

To test whether the choice of SED-fitting technique influences our conclusions based on the MCF measurements, we compare the MCFs obtained using SFRs based on \textsc{ProSpect} and \textsc{bagpipes} SED-fitting codes.
As described in Section~\ref{subsec:meth_MCF}, we use the rank of the property to weight the galaxies in the MCF measurement.
The left panel of Fig.~\ref{fig:Region_G23_SFR_comparision} shows the ratio of SFR ranks based on \textsc{bagpipes} to those based on \textsc{ProSpect}, plotted as a function of the \textsc{ProSpect}-based ranks.
The ranks diverge significantly toward the low-SFR end, suggesting that the two SED-fitting methods yield systematically different SFR estimates for galaxies with weak star formation.

The propagated effect on the MCFs is shown in the right panel of Fig.~\ref{fig:Region_G23_SFR_comparision}, where we compare the MCFs obtained using the SFR ranks from the two SED-fitting codes.
The MCFs differ significantly at small separation scales, where pairs are dominated by low-SFR galaxies and are therefore more sensitive to differences between the SED-fitting techniques.
However, the overall trend remains consistent, indicating that the choice of SED-fitting method does not alter our qualitative conclusions.

\begin{figure}
    \centering
    \includegraphics[width=\textwidth]{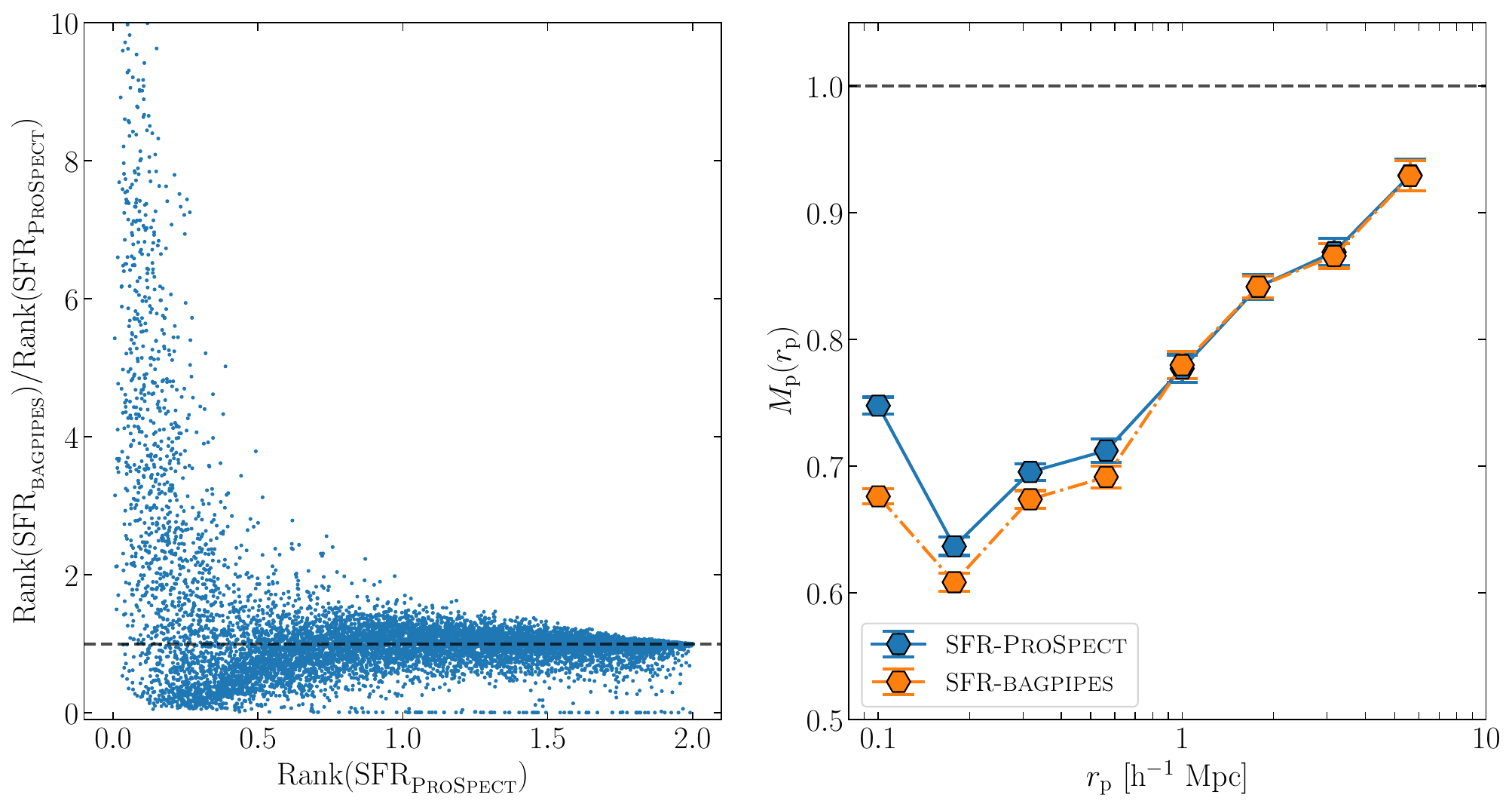}
    \caption{\textit{Left:} Ratio of ranks used for MCF computations using two different fitting codes: \textsc{ProSpect} and \textsc{bagpipes} vs rank used for SFR MCF of SFR MCFs of sample $G23$ estimated with two different fitting codes: \textsc{ProSpect} and \textsc{bagpipes}.}
    \label{fig:Region_G23_SFR_comparision}
\end{figure}

\bibliography{GAMA_MCF}{}
\bibliographystyle{aasjournalv7}



\end{document}